\documentclass[universe,review,accept,pdftex,moreauthors]    % <---- two or more authors
{Definitions/mdpi} 

\firstpage{1} 
\pubvolume{1}
\issuenum{1}
\articlenumber{0}
\pubyear{2024}
\copyrightyear{2024}
\datereceived{ } 
\daterevised{ } % Comment out if no revised date
\dateaccepted{ } 
\datepublished{ } 
\hreflink{https://doi.org/} % If needed use \linebreak
\Title{Future Perspectives for Gamma-ray Burst Detection from Space}

\TitleCitation{The future of GRBs from space}

\Author{Enrico Bozzo$^{1,\dagger}$*\orcidA{}, 
Lorenzo Amati$^{2}$\orcidI{},
Wayne Baumgartner$^{3}$\orcidb{},  
Tzu-Ching Chang$^{4}$\orcida{},   
Bertrand Cordier$^{5}$, 
Nicolas De Angelis$^{6,7}$\orcidB{},  
Akihiro Doi$^{8}$\orcidg{},
Marco Feroci$^{7}$\orcidW{},  
Cynthia Froning$^{9}$\orcidG{}, 
Jessica Gaskin$^{3}$\orcidL{}, 
Adam Goldstein$^{10}$\orcidC{}, 
Diego G\"otz$^{5}$\orcidM{}, 
Jon E. Grove$^{11}$\orcidi{}, 
Sylvain Guiriec$^{12}$\orcidJ{},  
Margarita Hernanz$^{13,14}$\orcidX{},
C.\ Michelle Hui$^{3}$\orcidH{}, 
Peter Jenke$^{15}$\orcidN{}, 
Daniel Kocevski$^{3}$\orcidh{},
Merlin Kole$^{6}$\orcidD{}, 
Chryssa Kouveliotou$^{12}$\orcidc{},  
Thomas Maccarone$^{16}$\orcidO{},
Mark L.\ McConnell$^{17,18}$\orcidP{}, 
Hideo Matsuhara$^{8}$, 
Paul O'Brien$^{19}$\orcidQ{}, 
Nicolas Produit$^{1}$\orcidE{},
Paul S.\ Ray$^{11}$\orcidR{},
Peter Roming$^{9}$\orcidS{}, 
Andrea Santangelo$^{20}$\orcidY{}, 
Michael Seiffert$^{4}$\orcidK{},  
Hui Sun$^{21}$\orcidT{},
Alexander van der Horst$^{12}$\orcidd{},  
Peter Veres$^{15}$\orcidF{},
Jianyan Wei$^{21}$\orcidf{},    
Nicholas White$^{12}$\orcidU{}, 
Colleen Wilson-Hodge$^{3}$\orcidU{}, 
Daisuke Yonetoku$^{22,8}$\orcide{},
Weimin Yuan$^{21,23}$\orcidV{}, 
Shuang-Nan Zhang$^{24,25}$\orcidZ{}. 	
}

\definecolor{arancio}{rgb}{1,0.5,0}
\definecolor{viola}{rgb}{0.7,0,1}
\definecolor{verde}{rgb}{0.2,0.7,0.7}
\definecolor{cobalt}{rgb}{0.0, 0.28, 0.67}
\definecolor{airforceblue}{rgb}{0.36, 0.54, 0.66}
\definecolor{ballblue}{rgb}{0.13, 0.67, 0.8}
\definecolor{battleshipgrey}{rgb}{0.52, 0.52, 0.51}
\definecolor{darkgreen}{rgb}{0.0, 0.2, 0.13}

\address{%
$^{1}$ \quad Department of Astronomy, University of Geneva, Chemin d'Ecogia 16, 1290, Versoix, Switzerland\\
$^{2}$ \quad INAF-OAS Bologna, via P. Gobetti 93/3, 40129 Bologna, Italy\\
$^{3}$ \quad NASA Marshall Space Flight Center, Huntsville, AL 35812, USA \\
$^{4}$ \quad Jet Propulsion Lab, 4800 Oak Grove Dr, Pasadena, CA 91109, USA \\
$^{5}$ \quad IRFU/D\'epartement d'Astrophysique,  CEA, Universit\'e  Paris-Saclay, F-91191, Gif-sur-Yvette, France\\
$^{6}$ \quad DPNC, University of Geneva, quai Ernest-Ansermet 24, 1205 Geneva, Switzerland; \\
$^{7}$ \quad INAF-IAPS Roma, via del Fosso del Cavaliere 100, I-00133 Roma, Italy \\
$^{8}$ \quad Institute of Space and Astronautical Science (ISAS), 3-1-1 Yoshinodai, Chuo-ku, Sagamihara, Kanagawa, 229-8510, Japan \\
$^{9}$ \quad Southwest Research Institute Space Science and Engineering Division, 6220 Culebra Road San Antonio, TX 78238\\
$^{10}$ \quad Science and Technology Institute, Universities Space Research Association, Huntsville, AL 35805, USA\\
$^{11}$ \quad Space Science Division, U.S. Naval Research Laboratory, Washington, DC 20375, USA\\
$^{12}$ \quad Department of Physics, George Washington University, Corcoran Hall, 725 21st Street NW, Washington DC 20052, USA \\
$^{13}$ \quad Institute of Space Sciences (ICE-CSIC), Carrer de can Magrans, s/n, Campus UAB, 08193 Bellaterra (Barcelona), Spain \\
$^{14}$ \quad Institut d’Estudis Espacials de Catalunya (IEEC), Barcelona, Spain \\
$^{15}$ \quad University of Alabama in Huntsville, Huntsville, AL 35805, USA\\
$^{16}$ \quad Department of Physics \& Astronomy, Texas Tech University Box 41051, Lubbock, TX 79409-1051, USA\\
$^{17}$ \quad Space Science Center, University of New Hampshire, Durham, NH 03824, USA  \\
$^{18}$ \quad Department of Earth, Oceans, and Space, Southwest Research Institute, Durham, NH 03824, USA \\
$^{19}$ \quad School of Physics and Astronomy, University of Leicester, Leicester LE1 7RH, UK\\
$^{20}$ \quad Institut f\"{u}r Astronomie und Astrophysik, Kepler Center for Astro and Particle Physics, Eberhard Karls, Universit\"{a}t, Sand 1, D-72076 T\"{u}bingen, Germany \\
$^{21}$ \quad National Astronomical Observatories, Chinese Academy of Sciences, Beijing 100101, China \\
$^{22}$ \quad College of Science and Engineering, School of Mathematics and Physics, Kanazawa University, Kakuma, Kanazawa, Ishikawa 920-1192, Japan \\
$^{23}$ \quad University of Chinese Academy of Sciences, Chinese Academy of Sciences, Beijing 100049, China \\
$^{24}$ \quad Key Laboratory of Particle Astrophysics, Institute of High Energy Physics, Chinese Academy of Sciences, 100049, Beijing, China \\
$^{25}$ \quad University of Chinese Academy of Sciences, Chinese Academy of Sciences, 100049, Beijing, China 
}

\firstnote{Most authors provided inputs on behalf of a larger collaboration.}
\corres{Correspondence: enrico.bozzo@unige.ch}

\abstract{Since their first discovery in the late 1960s, Gamma-ray bursts have attracted an exponentially growing interest from the international community due to their central role in the most highly debated open questions of the modern research of astronomy, astrophysics, cosmology, and fundamental physics. These range from the intimate nuclear composition of high density material within the core of ultra-dense neuron stars, to stellar evolution via the collapse of massive stars, the production and propagation of gravitational waves, as well as the exploration of the early Universe by unveiling first stars and galaxies (assessing also their evolution and cosmic re-ionization). GRBs have stimulated in the past $\sim$50~years the development of cutting-edge technological instruments for observations of high energy celestial sources from space, leading to the launch and successful operations of many different scientific missions (several of them still in data taking mode nowadays). In this review, we provide a brief description of the GRB-dedicated missions from space being designed and developed for the future. The list of these projects, not meant to be exhaustive, shall serve as a reference to interested readers to understand what is likely to come next to lead the further development of GRB research and associated phenomenology.}

\keyword{$\gamma$-ray astrophysics; Gamma-ray bursts; X-ray polarimetry; X-ray surveys; X-ray instrumentation} 

\begin{document}

\tableofcontents

\section{Introduction}

About 57 years have already gone by since the first discovery of a Gamma-ray burst (GRB) in 1967. This occurred genuinely by chance, while the \emph{Vela} military USA satellites were looking for evidence of nuclear bomb testing in space following the 1963 partial test-ban treaty (which prohibited testing in the atmosphere, outer space and underwater). The existence of GRBs was later revealed to the scientific community, stimulating a fast growing excitement for this fortunate discovery \citep{klebesadel}. A number of experiments were quickly developed and launched into space, mainly by the Soviet Union and the United States, to boost the detection of similar events and understand their true nature. The inevitable requirement of a space-based instrumentation to catch the unpredictable short and bright flash of high energy radiation from these events, largely contributed to boost the international race in the development of cutting-edge technological detectors with exponentially increasing sensitivity, as well as the provision of more and more advanced platforms to ensure their most proficient operations \citep[see, e.g.,][for recent hystorical reviews on GRBs]{2024VL}. 

A major leap forward in the GRB research was first provided by the Konus-WIND satellite \citep{konus,konus2} and the BATSE experiment on-board CGRO \citep{batse0,batse1}, which allowed the determination of the dichotomy into short and long GRBs, established the isotropic distribution of the events in the sky and led to the first characterization of their non-thermal emission \citep[see, e.g.,][and references therein]{batse2}. Subsequently,    
the Italian-Dutch {\em Beppo-SAX} satellite, launched in 1996, unveiled the cosmological nature of these events by identifying for the first time the 
GRB X-ray afterglows and providing arcsec-accurate localizations of their host galaxies \citep[enabling also multi-wavelength observations with ground-based observatories;][]{beppo2}. This was possible thanks to an on-board powerful suite of instruments covering a large energy-band (2--700~keV) and providing simultaneously a high sensitivity to faint sources via dedicated pointed observations and monitoring of a large fraction of the sky at once via large field-of-view (FoV) instruments \citep[achieving up to 20$\times$20~deg with full imaging and good localization capabilities of typically a few arcminutes;][]{beppo1}. 

The size of the GRB community and its efforts to probe and understand the physics of these uniquely powerful cosmological explosions have been growing ever since. GRBs are also widely recognized to have a key role in many of the most debated aspects of research in Astronomy and Astrophysics. These range from problems of fundamental physics, such as the equation of state of supra-density matter, to stellar evolution, via the collapse of massive stars and the production of both kilonovae and supernovae, to jet formation and dissipation, to the generation and propagation of gravitational waves, up to the quest for key Cosmological parameters \citep[see, e.g.][and references therein]{cosmology, cosmology2}. In the most recent years, GRBs have also gained renewed interest as key sources in the field of multi-messenger Astrophysics due to their association with Kilonovae and Gravitational Wave (GW) sources \citep[see the well known cases of GRB 170817A, GRB\,211211A and GRB\,230307A; e.g.,][and references therein]{gw,other,230307A}.

Following the decommissioning of \emph{Beppo-SAX} in 2002, a number of successful missions have provided dramatic advancements in the fields related to GRBs. Among these, AGILE \citep{2009A&A...502..995T}, Astro-SAT \citep{astrosat}, Fermi \citep{2009ApJ...697.1071A}, INTEGRAL \citep{1994ApJS...92..327W}, and the Neil Gehrels Swift Observatory \citep{2004ApJ...611.1005G}, are still operational and work in synergy also with ground-based multi-wavelength observatories to boost the detection and characterization of bright impulsive cosmological events (beside transients of different kinds). While an appraisal of the most relevant achievements of these missions can be found in recent reviews, the aim of the current paper is to provide interested readers an overview of the world-wide efforts from different teams to develop the GRB instruments of the future. These are expected to lead the scientific community toward the next leaps forward in the understanding of GRBs and associated phenomena. Different missions are presented in alphabetical order in the following sections. The list is not meant to be exhaustive and, given the heavily dynamic and fast-evolving nature of GRB research, it is likely (and desirable) that additional relevant missions set to provide important contributions to GRB research are being developed by other teams. 

%------------------------------------------------------------------------

\section{Einstein Probe}

\subsection{Mission Overview}

The Einstein Probe\footnote{\href{https://ep.bao.ac.cn}{https://ep.bao.ac.cn}} 
(EP) is a mission designed to monitor the sky in the soft X-ray band. It will perform systematic surveys and characterization of high-energy transients and monitoring of variable objects at unprecedented sensitivity and monitoring cadences. It has a large instantaneous FoV (3600 sq. deg.) that is achieved by using established technology of novel lobster-eye micro-pore optics (MPOs). EP also carries a conventional X-ray focusing telescope with a larger effective area to perform follow-up observations and precise localization of newly discovered transients. Public transient alerts will be issued rapidly to trigger multi-wavelength follow-up observations from the world-wide community. 

The primary science objectives of EP mission are: 
(1) Discover and characterize cosmic X-ray transients, particularly faint, distant and rare X-ray transients, in large numbers.
(2) Discover and characterize X-ray outbursts from otherwise normally dormant black holes.  
(3) Search for X-ray sources associated with gravitational-wave events and precisely locate them. 
EP is an international mission led by the Chinese Academy of Sciences (CAS) in collaboration with the European Space Agency (ESA), the Max Planck Institute for extraterrestrial Physics (MPE) in Germany and the Centre National d’Etudes Spatiales (CNES) in France. The mission was successfully launched on 2024 January 9 with a nominal lifetime of 3 years (5 years as a goal).

\begin{figure}
\centering
\includegraphics[width=3.7in]{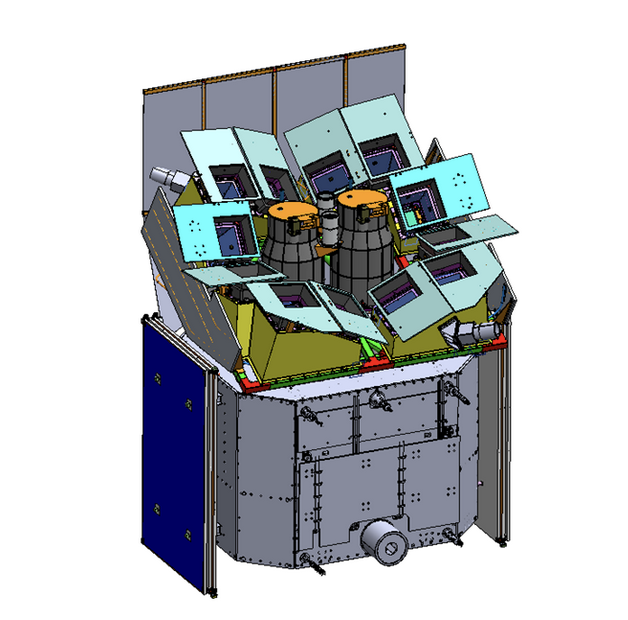}
\caption{Preliminary configuration of the EP payloads alignment, with twelve WXT modules surrounding two FXT telescopes.\label{fig:eppayload}}
\end{figure}

\subsection{Instrument Design}
 EP carries a Wide-field X-ray Telescope (WXT) with a large instantaneous FoV, which adopts a novel lobster-eye MPO technology. Complementary to this wide-field instrument is a Follow-up X-ray Telescope (FXT) with a large effective area and a narrow FoV. Figure \ref{fig:eppayload} shows the configuration of the EP payload.

\begin{itemize}
\item	Wide-field X-ray Telescope (WXT): There are 12 almost identical WXT modules, each with a FoV of $\sim$ 300 sq. deg. Each WXT module includes a lobster eye MPO mirror assembly with a focal length of 375\,mm, a focal plane detector array and an electronics unit. The mirror assembly of each module comprises 36 MPO optics mounted on a supporting alloy frame. 
An MPO optic is made of a thin plate with millions of square micro-pores perpendicular to the surface, slumped into a spherical shape. Incoming X-rays at a grazing-incidence angle are reflected off the walls of the pores and are brought onto a focal sphere with a radius of half the curvature of the optic. It produces true imaging with a characteristic cruciform point-spread function.
The detector array of each module is composed of four black-illuminated CMOS sensors, each of 6\,cm by 6\,cm in size and 4\,k by 4\,k in pixels. An aluminium layer of 200\,nm thickness  is coated on the surface to block incident optical and UV light. 
The nominal detection bandpass of WXT is 0.5--4 keV. 
\item	Follow-up X-ray Telescope (FXT): FXT is a set of two telescopes of the Wolter-I optics, which have almost the same design and are co-aligned. The design of the mirror assembly is similar to that of the eROSITA telescopes, which has a focal length of 1.6\,m. 
The focal plane detectors are built from pn-CCDs, and a set of thin and thick filters of aluminium layers are mounted on a filter wheel.  
\end{itemize}
\begin{figure}
\centering
\includegraphics[width=4.in]{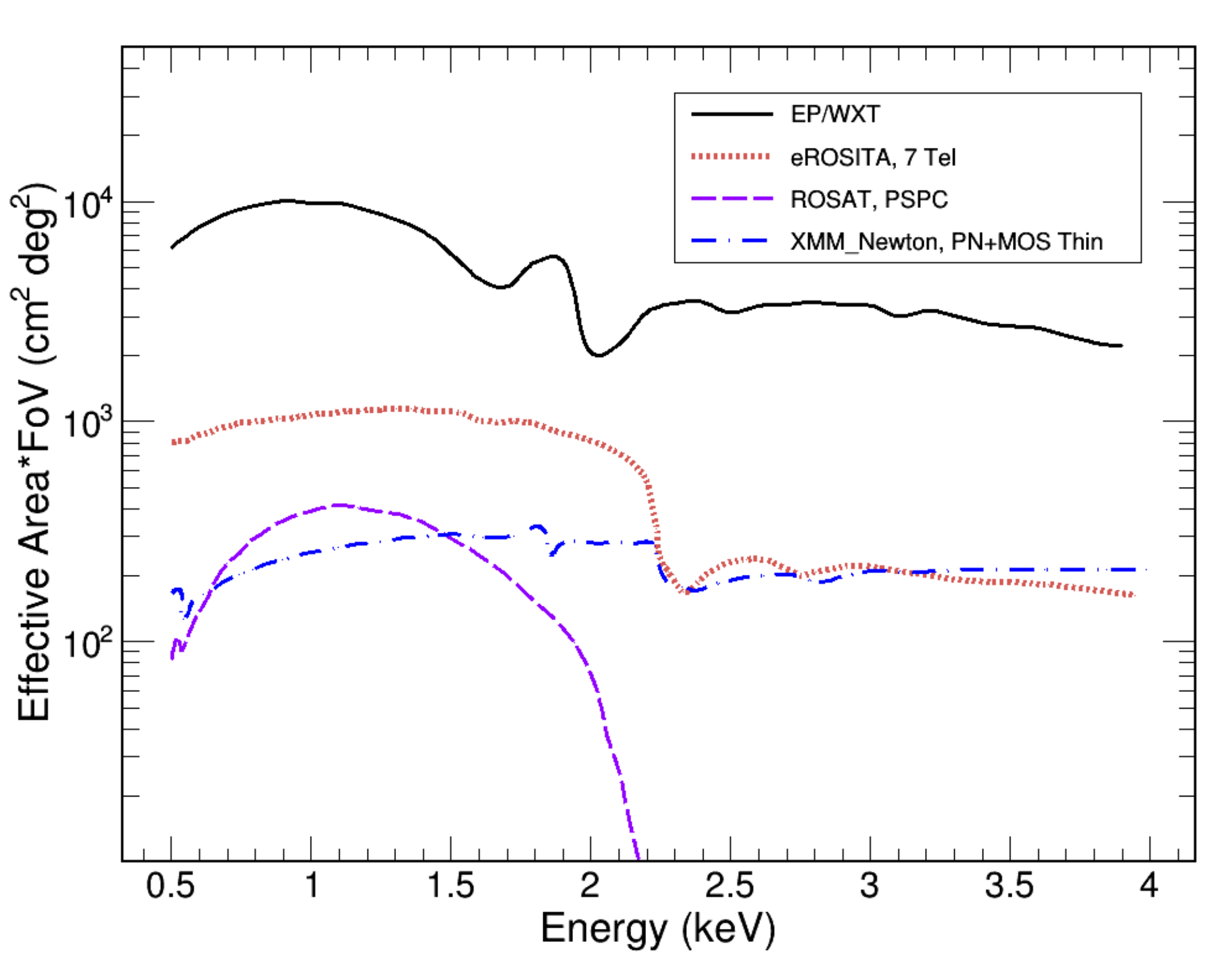}
\caption{Representative grasp (effective area times FoV) of WXT as a function of photon energy (black). As a comparison, the grasp parameters of several X-ray focusing telescopes are overplotted. Adapted from Figure 10 in \cite{Yuan2022}. \label{fig:epgrasp}}
\end{figure} 

\subsection{Expected Performance}

The angular resolution of WXT is about 5 arcmin (FWHM) for the central focal spot and the effective area in the range of 2--3 $\rm cm^2$. The grasp parameter of WXT is shown in Figure \ref{fig:epgrasp}. In a 1000 s exposure, a sensitivity of approximately (2--3)$\times10^{-11}$  $\rm ergs\,s^{-1}\,cm^{-2}$ in the 0.5--4 keV band can be achieved at the 5-$\sigma$ level. Such sensitivity and spatial resolution improve by one order of magnitude or more upon the previous and current wide-field X-ray monitors.

Operating in the 0.3--10 keV energy range, FXT has a narrow field of view (60 arcmin in diameter) and 
an effective area is about 300\,cm$^2$ at 1\,keV (for one unit).
The spatial resolution (PSF) is about 23\,arcsec in half-power diameter (HPD), which gives a source localization precision of 5--15 arcsec (90\,\% c.l.) depending on the source intensity.  
The FXT is responsible for the quick follow-up observations (within 5 minutes) of the triggered sources from WXT, and will also observe other interested targets as target of opportunity (ToO) observations.

Once a transient is detected with WXT, the spacecraft will slew to point FXT to the target for quick follow-up observations. Meanwhile the alert information of the transient will be down-linked quickly to the ground segment and made public to trigger follow-up observations. Quick command up-link for time-critical ToO observations is also possible. 

With the unprecedented capability of WXT and FXT, EP is expected to characterize the cosmic high-energy transients over wide time-scales and at high cadences, revealing new insights into a diverse set of systems including dormant black holes, neutron stars, supernova shock breakouts, active galactic nuclei, X-ray binaries, gamma-ray bursts, stellar coronal activity, and electromagnetic-wave sources and gravitational-wave events. In particular, EP will provide valuable data for the prompt emission of GRBs in the soft X-ray band, which has not been common in previous GRB detections. Meanwhile, EP will also monitor the variability of several types of X-ray sources in large samples all over the sky.

%------------------------------------------------------------------------

\section{eXTP}

\subsection{Mission Overview}

\begin{figure}[H]
\centering
\includegraphics[width=5.4in]{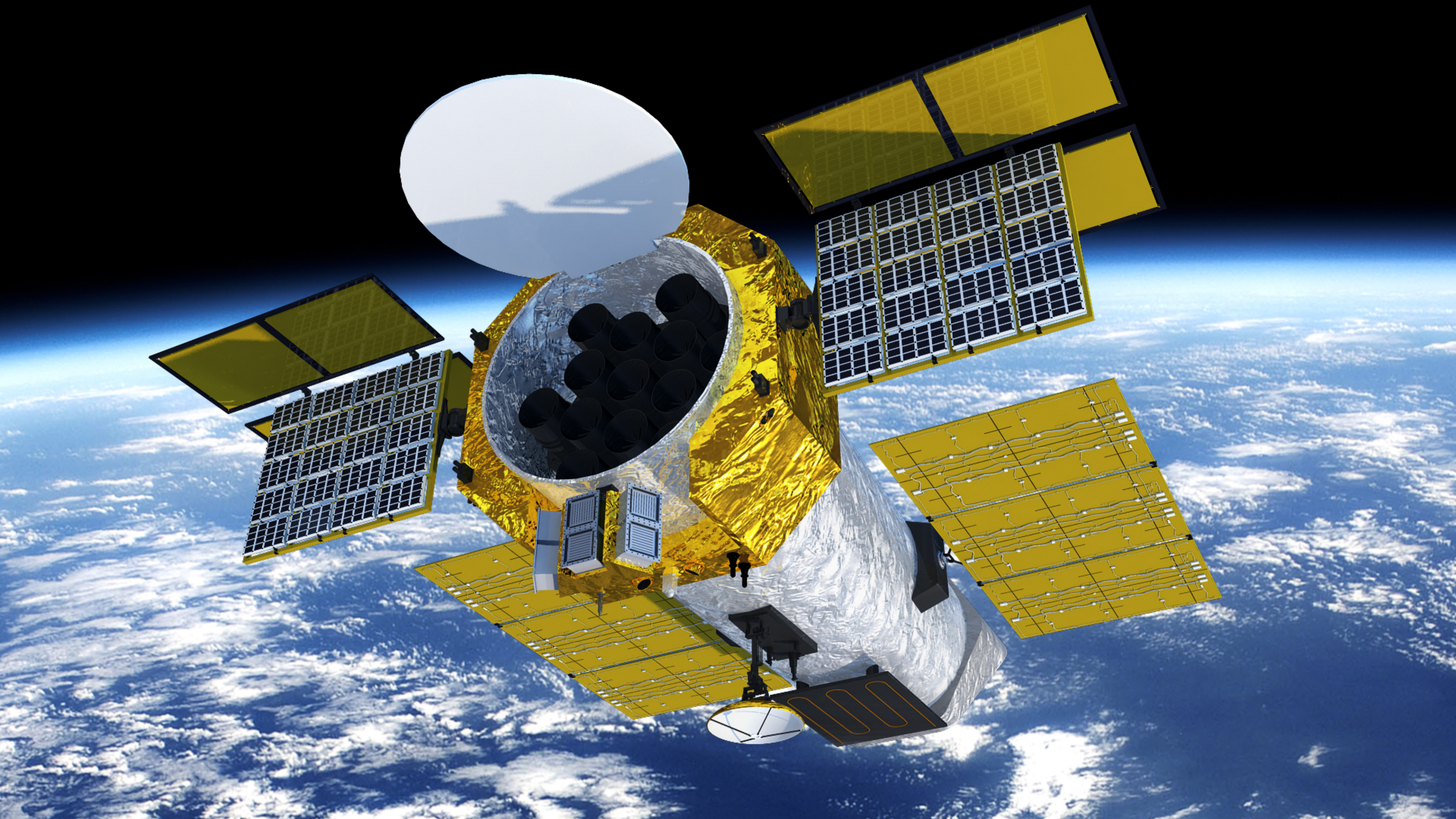}
\caption{An artist impression of the eXTP satellite. In the image we show the 13 nested shells optics of the SFA (9 telescopes) and PFA (4 telescopes), as well as the 6 cameras with coded masks composing the WFM. The LAD is placed on deployable panels equipped with solar shades. There are a total of 40 modules composing the LAD and each module hosts 16 SDDs. \label{fig:extp1}}
\end{figure}

The enhanced X-ray Timing and Polarimetry mission\footnote{\href{https://www.isdc.unige.ch/extp/}{https://www.isdc.unige.ch/extp/}} 
(eXTP) is designed to study mainly the state of matter under extreme conditions of density, gravity, and magnetic field \cite{extp}.The core science objectives of the mission are focused on the determination of the equation of state of matter at supra-nuclear density and the study of the dynamics of accretion/ejection flows under the influence of strong gravitational fields. Thanks to an extensive suite of innovative instruments, eXTP is designed also to be a general observatory for astrophysics, providing broad capabilities in the X-ray domain (0.5--50\,keV) to conduct timing, spectral, and polarimetric observations of a wide variety of Galactic and extra-Galactic sources. Although the mission is not focusing on GRB science, the availability of a large FoV and sensitive instrument, the WFM (see below), makes eXTP an important contributor to the possible detection and study of these events in the future. 

eXTP is planned to be the next flagship-class mission led by China and it is currently in an advanced design phase, where virtually all payload elements have reached a relatively mature technology and could be ready for implementation (see Figure~\ref{fig:extp1}). The mission is being studied by a large collaboration including a wide scientific community in China and many European member states. At present, a launch of opportunity has still to be identified, and programmatics are being cleared in order to possibly bring eXTP into space in 2028-2029.

\subsection{Instrument Design}

The eXTP scientific payload consists of four main instruments: the spectroscopic focusing array (SFA), the large area detector (LAD), the polarimetry focusing array (PFA), and the wide field monitor (WFM). The SFA comprises 9 identical Wolter-I grazing-
incidence X-ray telescopes, and is mainly used for spectral and timing observations of celestial sources in the energy range 0.5--10\,keV. In its current design, the SFA total eﬀective area is of 7400\,cm$^{2}$ (at 2\,keV) and the instrument FoV is circular with a diameter of 12\,arcmin. The instrument achieves an energy resolution of 180\,eV at 6\,keV and a timing resolution of 10\,$\mu$s in the whole operational energy band. The LAD is designed to perform photon-by-photon observations of X-ray sources in the 2--30~keV energy range. The LAD is exploiting the technology of innovative large-area Silicon Drift Detectors \citep[SDDs;][]{sdd} to provide an unprecedented large effective area, reaching 3.4\,m$^{2}$ at 8\,keV. The instrument is not designed for imaging purposes, and the FoV is limited to 1\,deg using micro-channel collimator plates \citep{mineo14} to simultaneously reduce source confusion and background. The LAD achieves an energy resolution of 260\,eV at 6~keV and a timing resolution of 10\,$\mu$s in the whole operating energy range. A similar instrument is also planned on-board the STROBE-X mission (see Sect.~\ref{sec:strobex}). The WFM exploits the same SDD technology coupled with coded masks to provide a coverage of about 5.5\,sr of the sky at any time in the energy range 2--50\,keV. The instrument is capable of localizing X-ray sources within an accuracy of $<$1\,arcmin and perform timing (spectral) investigations on these objects with a time (energy) resolution of 10\,$\mu$s ($<$300\,eV at 6\,keV). The WFM is equipped with an automatic alert system to promptly broadcast to the ground (within few tens of seconds at the most) the onset time and position of bright impulsive events detected on-board, including GRBs.  A similar instrument is also planned on-board the STROBE-X mission (see Sect.~\ref{sec:strobex}). The PFA provides polarimetric capabilities in the energy range 2--8\,keV, exploiting X-ray optics optimized for polarimetric observations and a similar detector technology compared to that currently flying on the IXPE mission \citep[the so-called gas pixel detectors, GPDs, detectors; see, e.g.,][and references therein]{ixpe}.  

\begin{figure}
\centering
\includegraphics[width=3.7in]{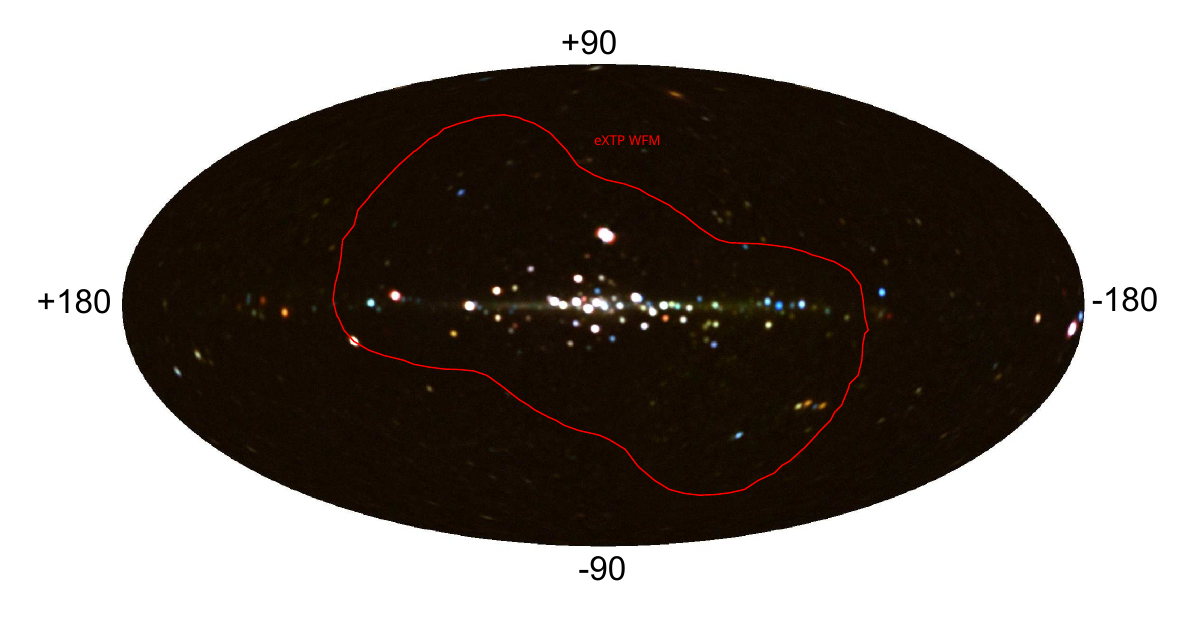}
\caption{The large FoV of the eXTP/WFM (red line) overlapped on a background map of the high energy sky that has been provided as a courtesy from the MAXI team (T.\ Mihara, RIKEN, JAXA). The red line corresponds to the total FoV covered by the WFM in the configuration used for eXTP as illustrated by  \citet{extpobs}. \label{fig:extp2}}
\end{figure}

\subsection{Expected Performance}

Given the capabilities of the eXTP/WFM (see Figure~\ref{fig:extp2}), a detection rate of about 100~GRB/year has been estimated. The WFM will be able to measure with good accuracy the spectral shape of the detected events and follow the hard to soft evolution of the prompt emission in the 2--50 keV band, where different GRB models are known to make different predictions \citep{zhang02, ghirlanda07}. This could provide information on the composition and magnetization of the emitting plasma, the geometry of the emission, and the structure of the jet and surrounding material. The combination of sensitivity and soft energy coverage of the WFM provides the opportunity to detect a few high redshift  (z$>6$) GRBs per year and many of the so-far elusive absorption features in tens of medium bright GRBs, probing (among others) the surroundings of GRB progenitors \citep[see, e.g.,][]{amati2000,gehrels09}. The WFM is also expected to detect each year up to 40 X-ray flashes \citep[see, e.g.,][]{Pelangeon2008}.

%------------------------------------------------------------------------ 
\section{Gamow}

\subsection{Mission Overview}

\begin{figure}
\centering
\includegraphics[width=5.in]{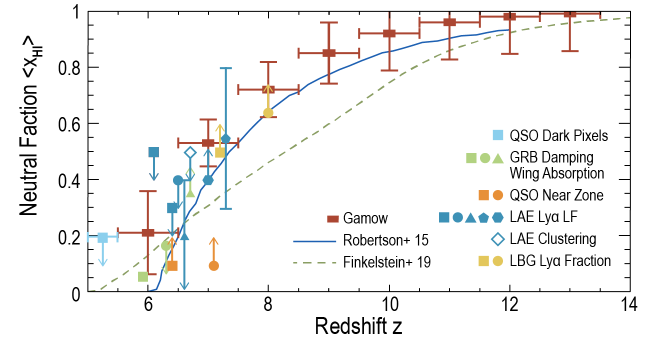}
\caption{Recent measurements of the neutral fraction $<$XHI$>$ versus redshift, z, by various techniques, including GRBs \citep[green; see][]{ota17}. The forecast Gamow results are shown as red \citep{lidz21}. The Fisher matrix forecast assumes 26 GRBs at z$>$5. We assume follow-up spectra with a signal-to-noise ratio (SNR) of 20 at the continuum per R=3000 resolution element and assume accurate host redshift determinations via metal absorption lines. The results illustrate that Gamow can map the re-ionization history in detail from z$\sim$6--14. Predicted re-ionization curves (blue and dashed red lines) illustrate the degree of variance between theoretical models compared to the Fisher forecasting. \label{fig:gamow1}}
\end{figure}

The Gamow Explorer mission \citep{gamow21} is optimized to: 1) probe the high redshift universe (z$>$6) when the first stars were born, galaxies formed and Hydrogen was re-ionized; and 2) enable multi-messenger astrophysics (MMA) by rapidly identifying electro-magnetic (EM) counterparts to gravitational wave (GW) events. GRBs have been detected out to z$\sim$9 \citep[see, e.g.,][and references therein]{salvaterra15,fryer22} and their afterglows are bright beacons lasting a few days that can be used to observe the spectral fingerprints of the host galaxy and intergalactic medium.  Gamow is designed to detect and rapidly identify high-redshift events. Rapid follow-up spectroscopy with the James Webb Space Telescope (JWST) and $>$6\,m ground-based telescopes provide NIR R$\sim2500$ spectra to determine the IGM neutral fraction versus redshift using the damping wing of the Lyman-$\alpha$ absorption line \citep[see Figure~\ref{fig:gamow1} from][]{lidz21}.

GRB afterglows are particularly advantageous for this measurement versus, e.g., using QSOs. GRBs have a featureless power-law spectrum, ideal for fitting the Ly-$\alpha$ absorption profile, are hosted by low mass galaxies directly tracing the typical ionization state of the IGM, and can be seen out to redshifts of 20, whereas the abundance of QSOs drops steeply with redshift. The same spectra will be used to determine metallicities from absorption lines to trace the early chemical enrichment and measure the escape fraction of ionizing photons that escape the galaxies to ionize atoms in the IGM. The co-moving GRB rate generally follows the star-formation rate (SFR). At higher redshifts, the GRB rate exceeds the SFR derived by other means. The greatly improved measurements of high-redshift GRB rate by Gamow, will provide crucial information to probe potential changes of the initial mass function of massive stars and star-formation rate at high redshift, and constrain GRB progenitors and their properties (e.g., luminosity distributions, progenitor binary fractions, etc.) in the high redshift Universe \citep{fryer22}. 

\subsection{Instrument Design}

The Lobster Eye X-ray Telescope (LEXT) with a wide FoV of $\sim$1350\,sq\,deg detects GRBs and locates them with arc-minute precision. The LEXT utilizes an array of slumped micro pore optics (MPOs) with 40\,$\mu$m square pores and a focal length of 30\,cm \citep{feldman21}. The focal plane uses heritage MIT-Lincoln Labs CCDs. The required performance is to detect at least 20 z$>$6 long GRBs over a 2.5\,year prime mission. A rapidly slewing spacecraft autonomously points the Photo-z Infra-Red Telescope (PIRT) within 100\,s to identify high redshift (z$>$6) GRBs. The photo-z technique takes advantage of the Hydrogen Lyman-$\sigma$ absorption which creates a sharp blue-ward drop out. A 30\,cm aluminum RC telescope passively cooled to 200\,K feeds a dichroic prism beam splitter to place five images onto a single H2RG detector covering the 0.6 to 2.5\,$\mu$m band \citep{seiffert21}. The FoV is 10~arcmin square. This design provides the required 15 $\mu$Jy (21 mag AB) 5-$\sigma$ sensitivity in a single 500\,s exposure \citep[see also][]{pirt2}.  
The high redshift science objectives require rapid follow-up observations to provide NIR medium resolution spectroscopy (R$\sim$3000) to measure the profile of the Ly-$\sigma$ absorption line and metal absorption lines from the host galaxy. To obtain sufficient signal to noise in a reasonable observing time requires 6\,m class or greater telescopes. Ground-based telescopes begin observations within an hour and JWST and the ELT within 2--3 days (Kann et al.\ 2024, to be submitted). An L2 orbit provides $>$95\,\% observing efficiency with pointing optimized for these follow-up observatories. A low bandwidth continuous S-band connection provides real time alerts to the ground. 
\begin{figure}
\centering
\includegraphics[width=3.7in]{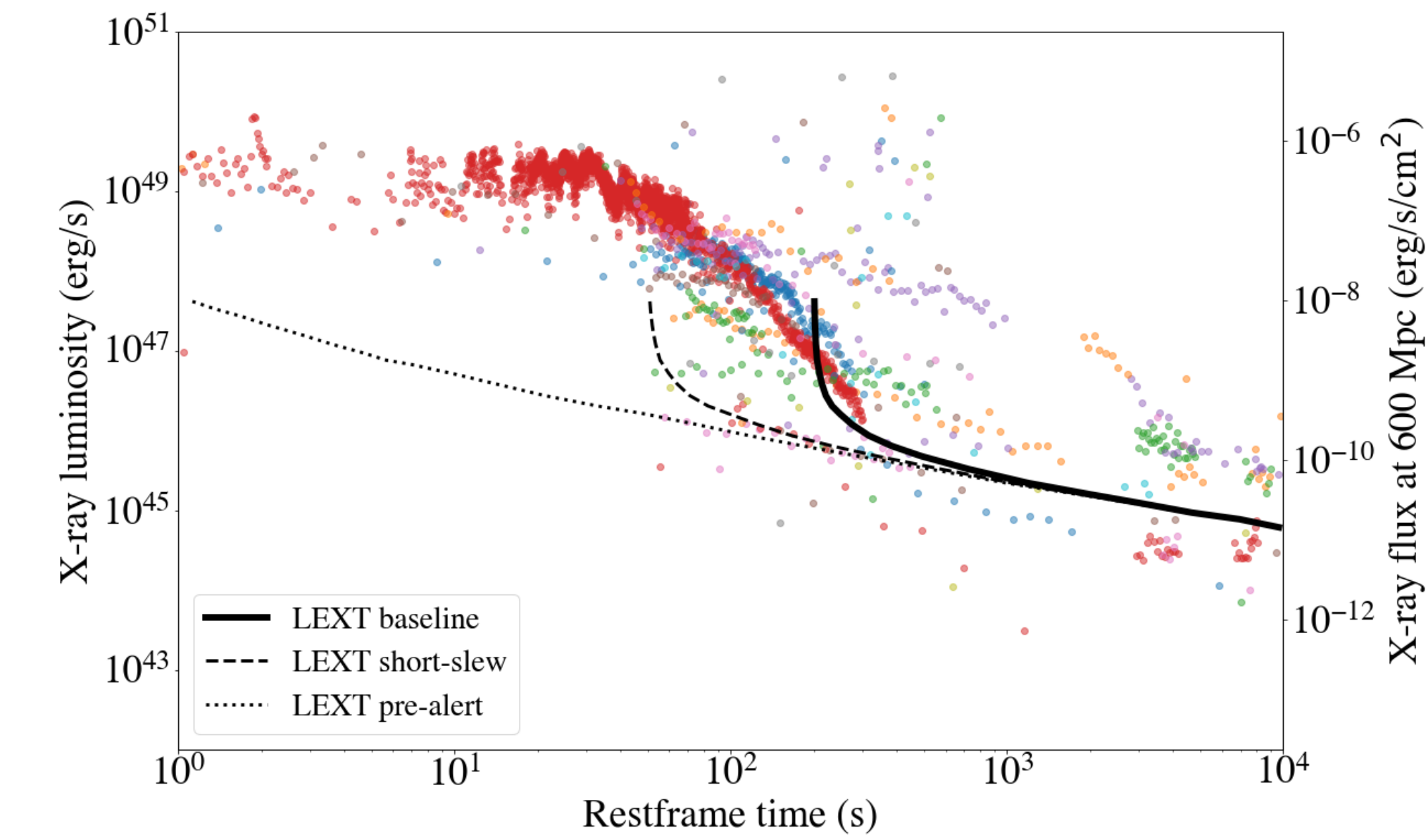}
\caption{The ability of Gamow LEXT to detect the afterglows of on-axis BNS and NSBH mergers. X-ray luminosity (left axis) and X-ray flux at 600\,Mpc (right axis) for Swift short GRBs at known redshift. The black line is the LEXT sensitivity starting at 0\,s (base-line). In practice, it will frequently be earlier than this (short-slew), and sometimes before the BNS merger (pre-alert; figure from \cite{gamow21} and see \citet{chan,nitz21,Banerjee23}). \label{fig:gamow2}}
\end{figure} 

\subsection{Expected Performance}

The Gamow capabilities are also optimized for the identification of EM counterparts to binary neutron star (BNS) and neutron star black hole (NSBH) mergers detected by the A+ generation of GW detectors. Within $<$200\,s of a GW alert the real time low bandwidth link uploads commands to re-point the LEXT at the GW uncertainty region so as to detect and locate to 3\,arcmin precision the accompanying short GRB afterglow. Figure 2 from \citet{gamow21} shows measurement predictions are based on extrapolation of Swift short GRB afterglow detections to the LEXT pass-band and 600\,Mpc source distances (the horizon of the LVK A+ GW detectors). Approximately 80\,\% of Swift short GRBs have X-ray afterglow detections. These demonstrate that they will be detectable by Gamow at BNS-appropriate distances. If a transient X-ray source is detected, an autonomous PIRT pointing will refine the position to 1\,arcsec precision and use simultaneous five-band photometry to follow the early phase of the merger. The X-ray and Optical-NIR multi-band and multi-wavelength capabilities combined with an agile spacecraft will enable a broad swath of time domain astronomy science, very similar to Swift \citep{gehrelscannizzo2015}. While waiting for GRBs and GW events we envision Gamow will undertake a community driven comprehensive Time Domain Astronomy program. The Gamow Explorer was proposed to the 2021 NASA MIDEX call. While it was not successful, there are considerations to re-propose to future NASA opportunities.

%------------------------------------------------------------------------
% hi-z gundam

\begin{figure}[t]
\centering
\includegraphics[width=4.8in]{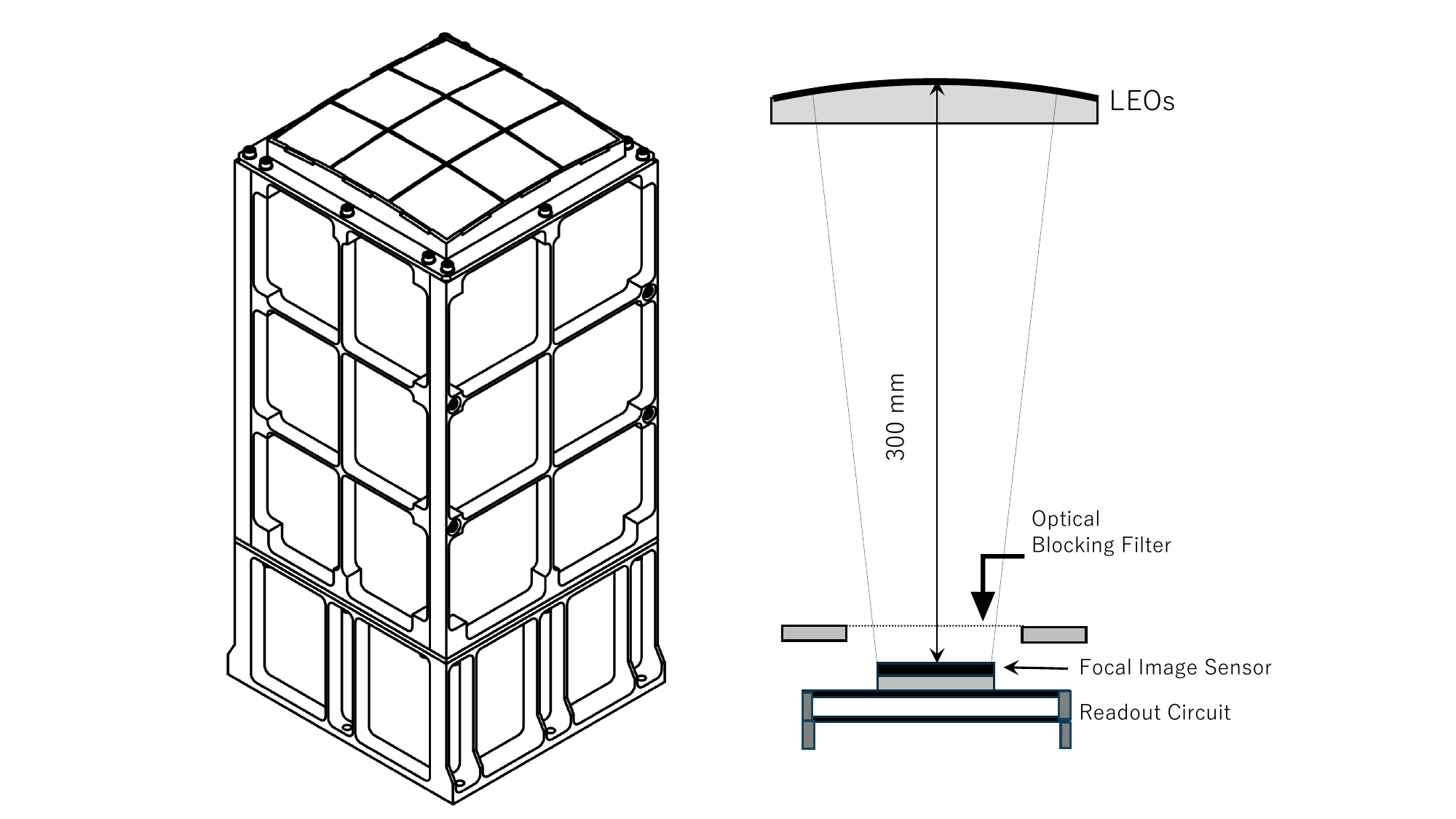}
\includegraphics[width=2.7in]{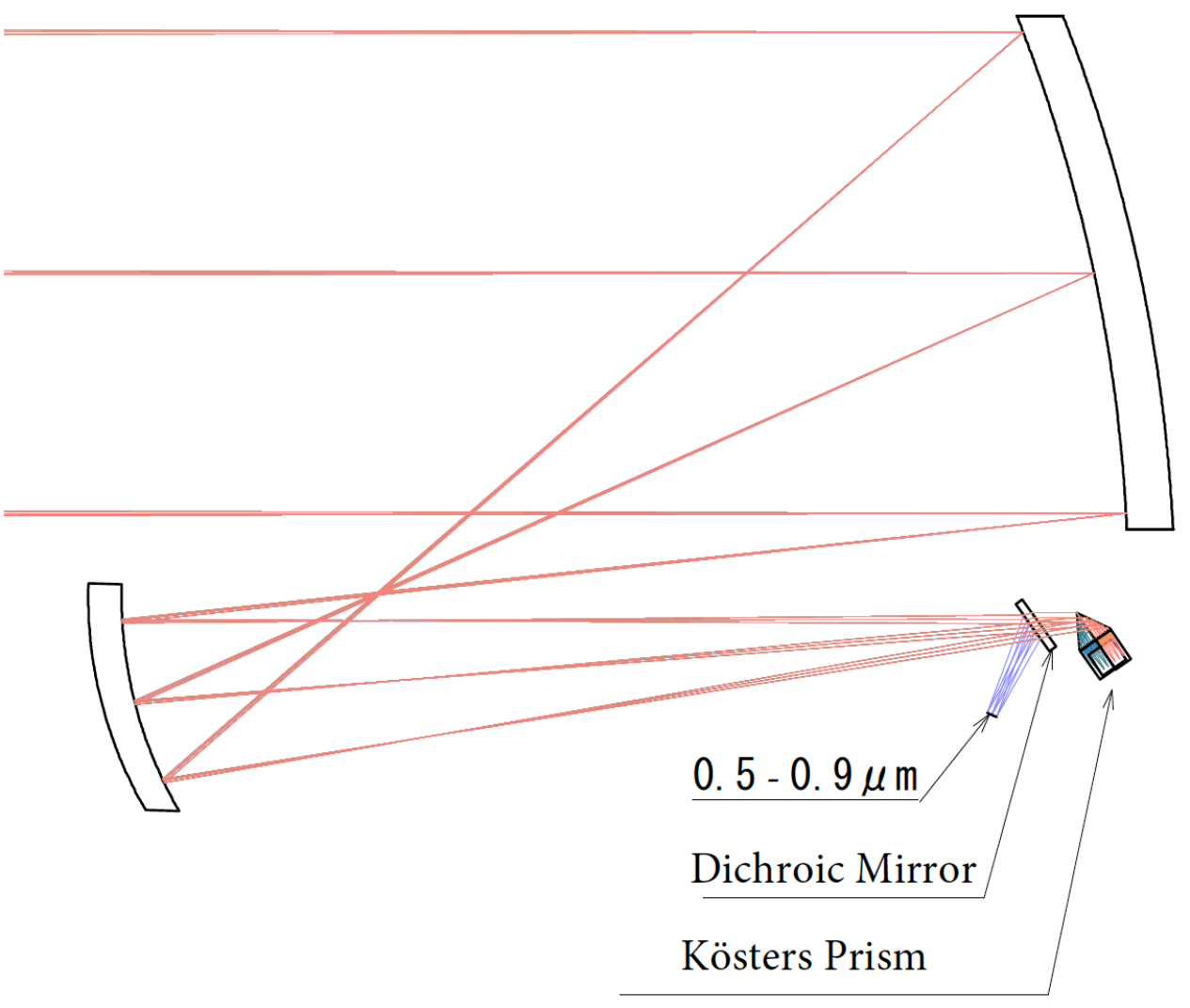}
\includegraphics[width=2.7in]{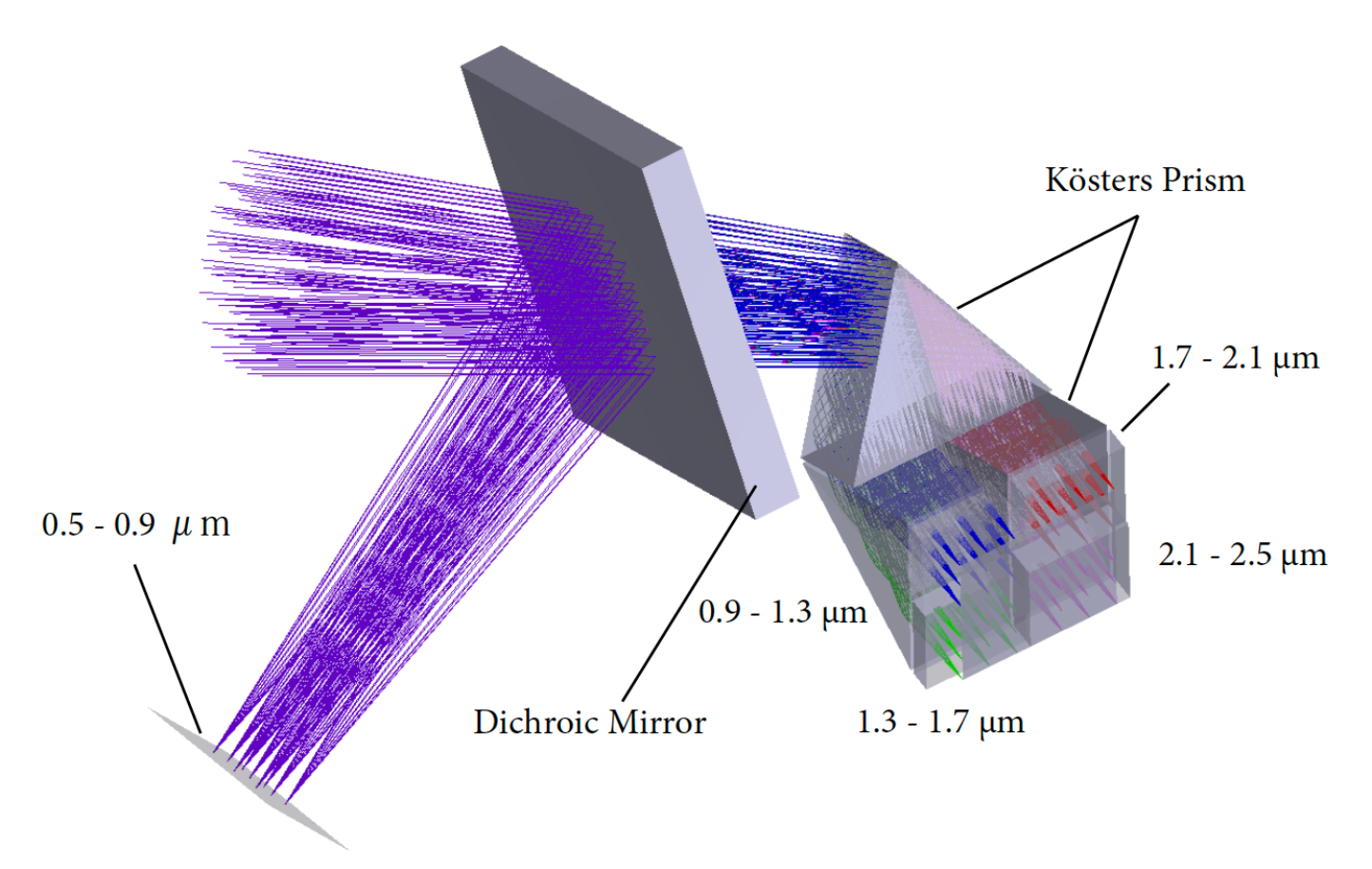}
\caption{\textbf{a}) A schematic view of the HiZ-GUNDAM WFXM (top; copyright MEISEI ELECTRIC Co., LTD.). \textbf{b}) Schematic views of the HiZ-GUNDAM NIRT (middle and bottom).}
\label{HZG_payloads}
\end{figure}

\section{HiZ-GUNDAM}

\subsection{Mission Overview}

The High-z gamma-ray bursts for unraveling the dark ages mission, HiZ-GUNDAM\footnote{\href{https://www.isas.jaxa.jp/en/missions/spacecraft/future/hiz-gundam.html}{https://www.isas.jaxa.jp/en/missions/spacecraft/future/hiz-gundam.html}}, is a future satellite mission, a competitive medium-class mission in ISAS/JAXA, designed to advance time-domain astronomy through the observation of high-energy transient phenomena \cite{yonetoku2020, yonetoku2022}. Two scientific goals are defined: (1) the exploration of the early universe with high-redshift gamma-ray bursts and (2) contribution to multi-messenger astronomy. These scientific objectives require observational capabilities to detect high-energy transients and to carry out automatic/rapid follow-ups in the near-infrared band. This Section provides the description of the current specifications and designs of the satellite and mission payloads, but it is important to note that they are subject to change in subsequent studies and developments.

\subsection{Instrument Design}

HiZ-GUNDAM has two types of mission payloads: the wide-field X-ray monitor (WFXM) and the near-infrared telescope (NIRT). These payloads are optimized and minimized to achieve the scientific goals \cite{li2020, ogino2020, sawano2020, goto2022, tsumura2020}. We show a schematic view in Figure~\ref{HZG_payloads}, and basic specification of WFXM and NIRT in Table~\ref{tab:HZG_WFXM} and \ref{tab:HZG_NIRT}, respectively.

The wide-field X-ray monitors (WFXM) consist of Lobster Eye optics arrays and focal imaging sensors (pnCCD). They are designed to detect high-energy transients within a wide field of view of $> 0.5~\mathrm{steradian}$ in the 0.5--4\,keV energy range. The current design of a single module of WFXM includes a lobster-eye optic array of $3 \times 3$ with a 300\,mm focal length and a pnCCD with a pixel size of approximately 100\,$\mu$m on a $55 \times 55$\,mm$^2$ format at the focus. Multiple modules are installed aboard the spacecraft to monitor the wide FoV.

The near-infrared telescope (NIRT) has an aperture size of 30\,cm in diameter and simultaneously observes at five wavelength bands between 0.5--2.5\,$\mu$m using a dichroic mirror and K\"{o}sters. The NIR telescope is cooled to $< 200$\,K by radiative cooling to maintain the best sensitivity up to the 2.5\,$\mu$m band. The mirrors and structure of NIRT are primarily made of aluminum alloy to ensure imaging performance remains consistent even if the temperature of the telescope changes in orbit, i.e., an athermal configuration.

\begin{table}
\centering
\caption{HiZ-GUNDAM Wide Field X-ray Monitor.}
\label{tab:HZG_WFXM}
\begin{tabular}{ l c } 
\hline
Item & Specification \\
\hline
Optics & Lobster Eye\\
Focal Length  & 300\,mm \\ 
Focal Detector & pnCCD \\
Size of Focal Detector  & $55 \times55$\,mm${^2}$ \\
Energy Band  & 0.5--4.0\,keV\\
Field of View  & $> 0.5$ str (in total)\\
Sensitivity (100\,s) & $\sim 10^{-10}$ erg cm$^{-2}$ s$^{-1}$ \\
Localization Accuracy & $\sim 3$\,arcmin\\ 
Time Resolution & $\sim 0.1$\,s\\
\hline
\end{tabular} 
\end{table}

\begin{table}
\centering
\caption{HiZ-GUNDAM Near Infrared Telescope.}
\label{tab:HZG_NIRT}
\begin{tabular}{ l c } 
\hline
Item & Specification \\
\hline
Telescope Type  & Offset/Athermal \\
Aperture Size & 30\,cm\\
Telescope Temp. & $< 200$\,K \\ 
Focal Detector & HyViSi (Optical) \\
                         & HgCdTe (NIR)\\
Field of View & $15 \times 15$\,arcmin$^2$ \\ 
\hline
Band & Sensitivity (2~min$\times$5) \\
0.5--0.9~$\mu$\,m & 21.4~mag(AB)\\
0.9--1.3~$\mu$\,m  & 21.3~mag(AB)\\
1.3--1.7~$\mu$\,m  & 21.4~mag(AB)\\
1.7--2.1~$\mu$\,m  & 20.8~mag(AB)\\
2.1--2.5~$\mu$\,m  & 20.7~mag(AB)\\
\hline
\end{tabular}
\end{table}

\subsection{Expected Performance}

\begin{figure}
\centering
\includegraphics[width=12cm,pagebox=cropbox,clip]{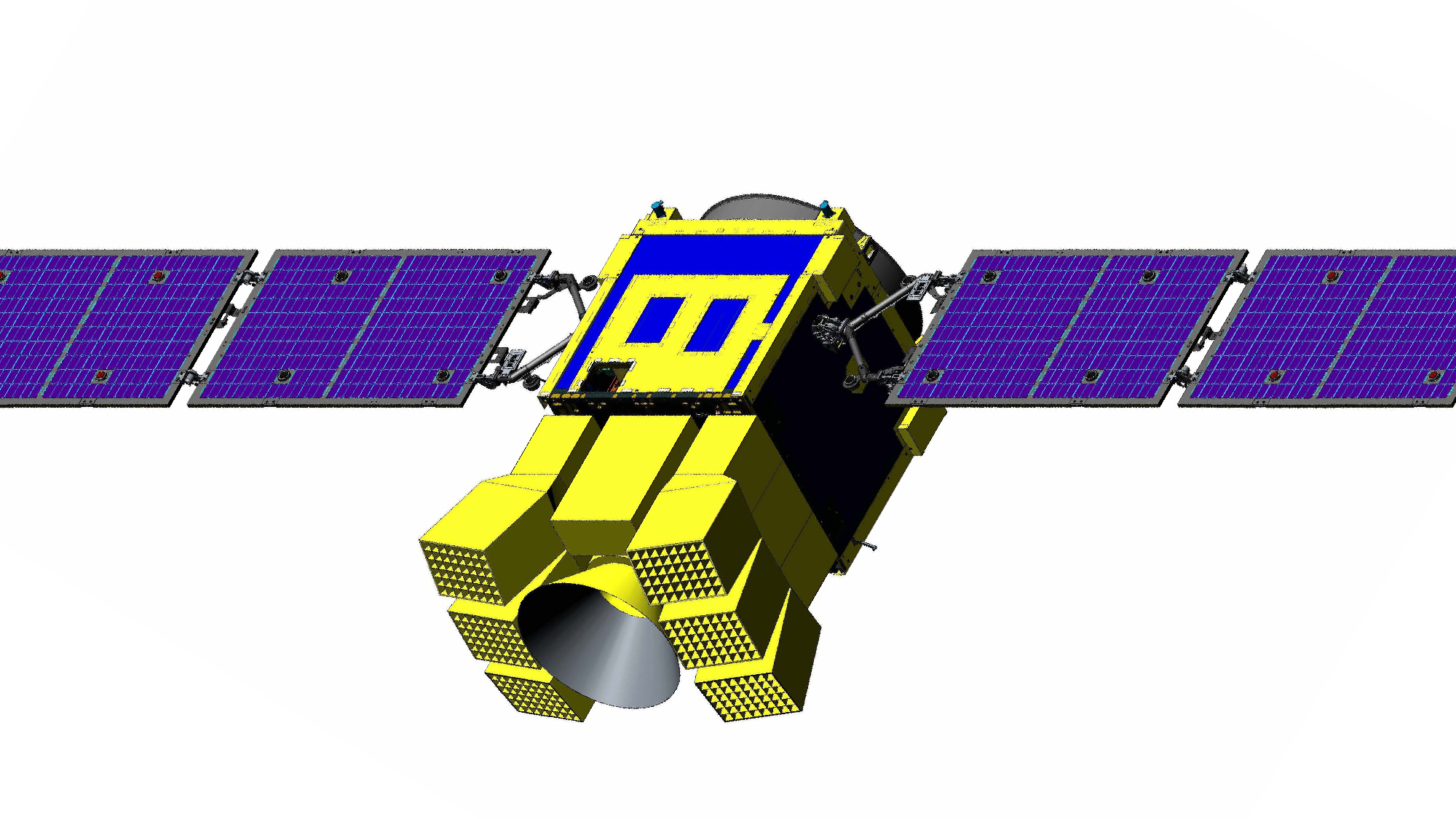}
\caption{Schematic view of the HiZ-GUNDAM satellite.}
\label{fig:HZG_satellite}
\end{figure}

In Figure~\ref{fig:HZG_satellite}, we show schematic views of HiZ-GUNDAM satellite. The size of satellite bus is about 1\,m$^3$ except for the solar paddles and mission payloads. For orbital placement, we have chosen a sun-synchronous polar orbit along the twilight line. This choice is based on the favorable thermal conditions for the NIRT, 
which is cooled down to $< 200$\,K. However, X-ray observation may face disturbances from both South Atlantic Anomaly and aurora belt at high latitude polar region. The nominal operation of HiZ-GUNDAM follows the sequence outlined below:
\begin{enumerate}
\item Set the satellite attitude to a solar separation angle of 120\,degrees and a forward movement of 50 degrees.
\item Maintain the inertial pointing direction for approximately 560\,s, during which HiZ-GUNDAM monitors X-ray transients.
\item After the monitoring time, the satellite slews to the same attitude configuration as described in (1), 
but with a different pointing direction to prevent thermal radiation from the Earth exposing the NIRT.
\end{enumerate}
This nominal sequence is repeated, and the observation of eight different fields of view are performed in every orbit. The satellite continuously monitors the X-ray sky, except during the maneuver. This optimized sequence facilitates follow-up observations with the NIR telescope. HiZ-GUNDAM can execute this operation for 97\,\% of the GRBs discovered by itself, ensuring more than 10 minutes of follow-up time in each case.

%------------------------------------------------------------------------

\section{LEAP}

\begin{figure}
\centering
\includegraphics[width=3.7in]{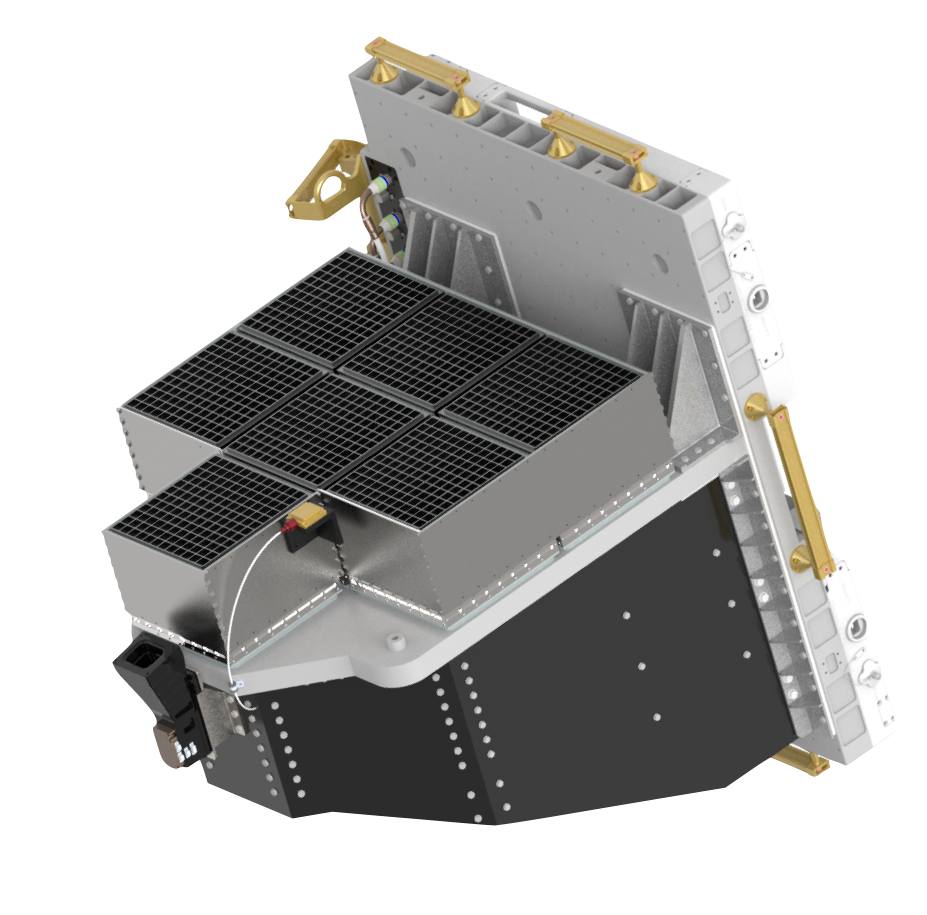}
\caption{The LEAP payload includes seven independent polarimeter modules. When mounted on the outside of the ISS, the array of modules points toward the local zenith and scans the sky for GRBs. \label{fig:leap}}
\end{figure}

\subsection{Mission Overview}

The LargE Area burst Polarimeter\footnote{\href{https://sti.usra.edu/in-the-news/leap-mission-study/}{https://sti.usra.edu/in-the-news/leap-mission-study/}} 
(LEAP) is a mission that was proposed in 2021 to NASA's Astrophysics Mission of Opportunity Program. The LEAP instrument is a wide FoV Compton polarimeter that measures GRB polarization over the energy range from 50\,keV to 1\,MeV and performs GRB spectroscopy from 20 
\,keV to 5\,MeV.  If approved (a final decision on the selection is expected during the first quarter of 2024), it will be deployed as an external payload on the International Space Station (ISS) in 2027 for a three year mission \cite{McConnell.2021,Onate-Melecio.2021}.  The LEAP science investigation is based on the ability to distinguish between three classes of GRB models \cite{Toma.20097gfh}. The baseline science investigation requires the observation of 65 GRBs with a minimum detectable polarization (MDP) of 30\,\% or better. Evidence of polarized $\gamma$-ray emission in GRBs ($>$ 100\,keV) has been accumulated in recent years, but the limited sensitivity of these measurements does not yet yield a clear picture of the underlying physics \cite{McConnell.2017,Tatischeff.2019,Chattopadhyay.2021ny}. A sensitive and systematic study of GRB polarization, such as the one that will be provided by LEAP, is needed to remedy this situation.

\subsection{Detector Design}

The LEAP payload (Figure~\ref{fig:leap}) consists of an array of seven independent polarimeter modules, each with a 12$\times$12 array of optically isolated high-Z and low-Z scintillation detectors read out by individual PMTs. Each polarimeter module includes 84 plastic scintillator elements, 58 CsI(Tl) scintillator elements, and two $^{60}$Co calibration sources. Each calibration source consists of a small plastic scintillator doped with $^{60}$Co, which permits electronic tagging of each decay via the coincident $\beta$ particle. As a Compton polarimeter, scatter events recorded by the scintillator array within each module are used to measure polarization. Within each module, the arrangement of plastic and CsI(Tl) scintillation detectors is designed to optimize the polarization response. The dominant type of scatter event is one involving only two detector elements, in which incident photons scatter from a low-Z  plastic detector element into a high-Z CsI(Tl) element. The distribution of azimuthal scatter angles for these events provides a polarization signature.  The total effective area for polarimetry is $\sim1000$ cm$^2$ at energies above 100\,keV.  Since the total energy deposit is a sum of the energy deposits in all triggered elements, spectroscopic information is provided by all types of events of any multiplicity (both singles event and scatter events).  To characterize the GRB parameters, spectroscopic measurements (20-5000\,keV) are obtained using all event types (both multiple and singles events), with a total effective area that reaches $>3000$ cm$^2$ between 50 and 500\,keV. 

To accurately reconstruct the source spectrum and polarization, LEAP self-sufficiently determines the source direction using singles events from all 420 CsI(Tl) elements, whose relative response provides the source localization. Localization errors of $< 5^{\circ}$ (1$\sigma$) are obtained at a rate of about 40 per year. This is sufficiently precise to enable rapid followup by many ground-based instruments using the rapid burst response messages that will be generated and distributed (in real time) by LEAP.

\subsection{Expected Performance}

\begin{figure}
\centering
\includegraphics[width=3.8in]{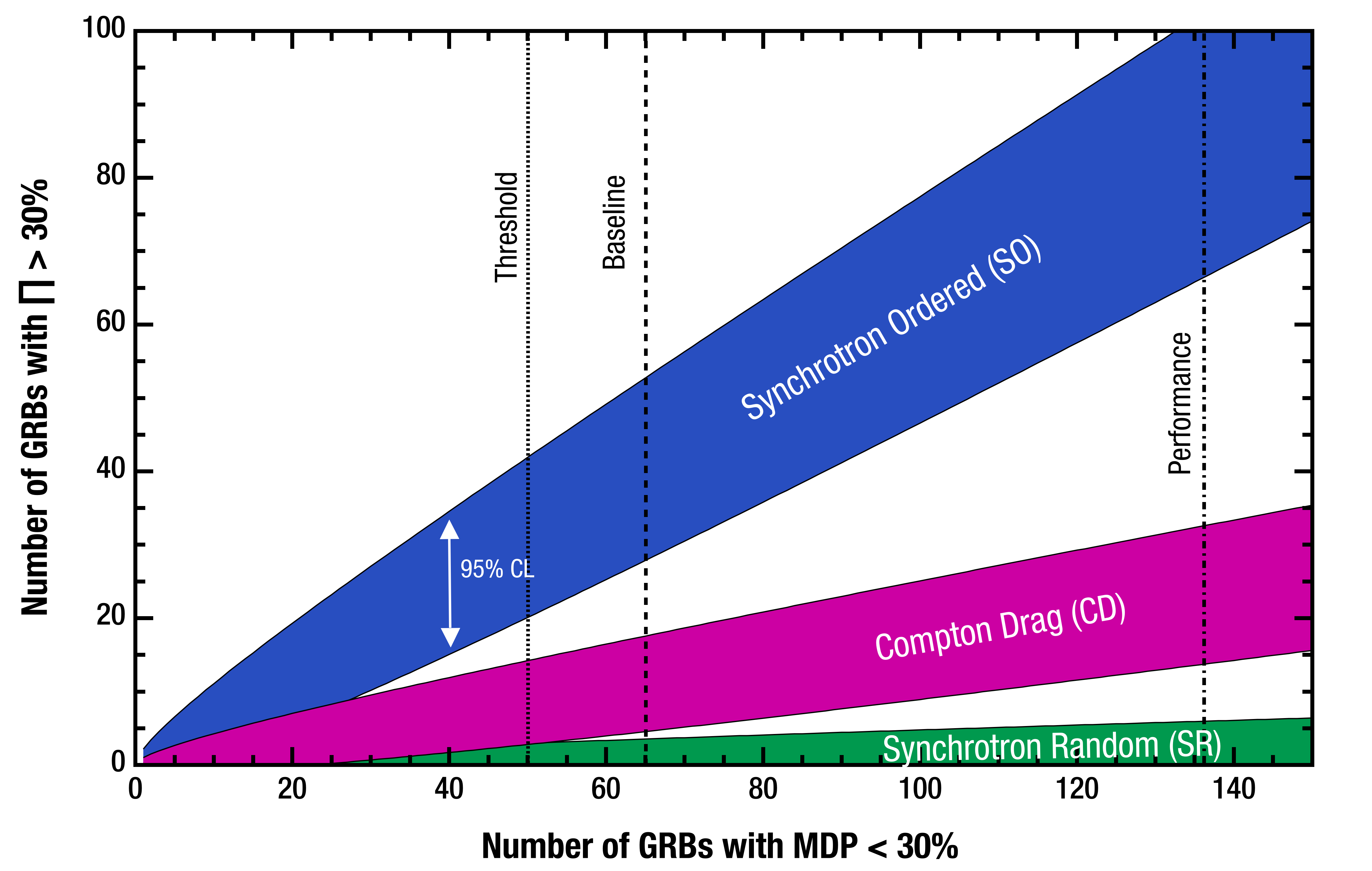}
\caption{The number of GRBs measured with a polarization degree $> 30\%$ versus the number of GRBs with an MDP $< 30\%$ provides a convenient way to distinguish between different model classes. For the case shown here, 65 GRB measurements are required to distinguish the three model classes. LEAP will be able to measure a total of 135 GRBs with MDP $< 30\%$ (and 23 GRBs with MDP $< 10\%$) during its baseline 32 months of science operations.\label{fig:leap-performance}}
\end{figure}

As a wide FoV instrument, LEAP maintains some level of polarization sensitivity out to at least 75$^{\circ}$ off-axis, providing an effective FoV of $\sim$1.5$\pi$\,sr.  Minimal obstructions from ISS within the FOV maximize sky exposure and minimize photon absorption and scattering effects. However, scattered flux from ISS structures (such as solar panels) must always be considered in the analysis of both the spectrum and the polarization.  
Response simulations spanning a range in energy, spectral shape, and incidence direction have been used in estimates of instrument performance. Figure~\ref{fig:leap-performance} shows, for a three year mission, the number of expected GRBs as a function of MDP. These estimates show that LEAP will attain its requirement of 65 GRBs with $<30$\,\% MDP, with significant margin. 

%------------------------------------------------------------------------

\begin{figure}
\centering
\includegraphics[width=4.5in]{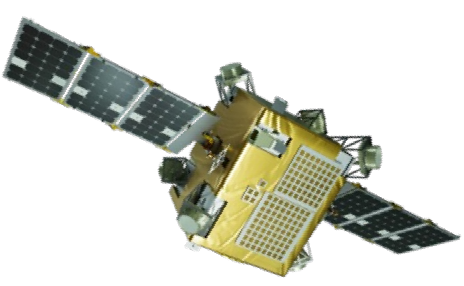}
\caption{Six scintillating detectors positioned for an instantaneous all-sky field of view, no slewing required. Coupled with a cislunar orbit, MoonBEAM provides an unprecedented all-sky sensitivity that cannot be achieved in Low Earth Orbit.\label{fig:moonbeam}}
\end{figure}

\begin{figure}[H]
\centering
\includegraphics[width=5in]{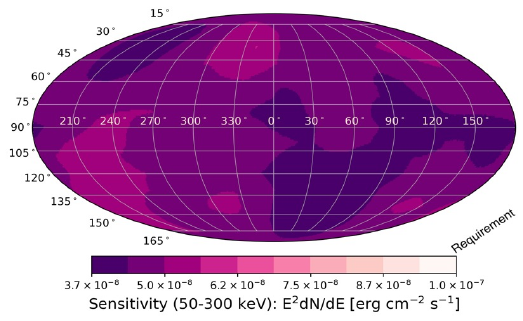}
\caption{MoonBEAM’s 1-second limiting flux sensitivity across the entire sky in any instant, providing the continuous sensitivity needed to study astrophysical jet formation and emissions.\label{fig:MB-allsky}}
\end{figure}

\section{MoonBEAM}

\subsection{Mission Overview}

Moon Burst Energetics All-sky Monitor (MoonBEAM) is a proposed gamma-ray mission to observe the entire sky instantaneously for relativistic astrophysical explosions from a cislunar orbit. It is designed to explore the behavior of matter and energy under extreme conditions by observing the prompt emission from GRBs, identifying the conditions capable of launching transient relativistic jets and the origins of high-energy radiation from the relativistic outflows. MoonBEAM provides the essential continuous all-sky gamma-ray observations for time-domain and multi-messenger astrophysics by reporting on any prompt emission of a GRB and by providing the critical first alerts to the community for contemporaneous and follow-up observations.
 
\subsection{Detector Design}

MoonBEAM achieves instantaneous, all-sky coverage by positioning six gamma-ray detector assemblies at the corners of the spacecraft to minimize blockage (see Figure~\ref{fig:moonbeam}), and by deploying the observatory in an Earth-Moon Lagrange Point 3 cislunar orbit instead of Low Earth Orbit (LEO) to reduce particle background from radiation belts, atmospheric interactions, and planetary occultation from 30\,\% to $<< 1$\,\% of the sky. 

Each detector assembly consists of a NaI(Tl)/CsI(Na) phoswich scintillator coupled to flat panel photomultiplier tubes. The phoswich design allows for both localization improvement and increased effective area for spectroscopy. It is sensitive to 10--5000\,keV photons, with an energy resolution better than 12\,\% at 662\,keV.

\subsection{Expected Performance}

MoonBEAM is expected to detect more than 1000 GRBs over 30 months of operation, with an intrinsic localization capability similar to that of the \textit{Fermi} Gamma-ray Burst Monitor. In the absence of a detection, MoonBEAM provides unprecedented sensitive gamma-ray upper limits for any externally detected transients such as mergers seen in gravitational waves and supernovae detected in optical wavelengths, estimated to be in the order of thousands over the operation time of MoonBEAM. Figure~\ref{fig:MB-allsky} shows the 1-second limiting flux sensitivity of MoonBEAM across the entire sky. 

The wide-field, sensitive, and continuous gamma-ray coverage is necessary to advance our current understanding of astrophysical jet formation, structure, and evolution. MoonBEAM achieves sensitivity improvement over current missions in LEO because of the combined advantage of its cislunar orbit and instrument design. It will join the Interplanetary Gamma-Ray Burst Timing Network \cite{MB:Hurley2013} as one of the few missions outside of LEO with gamma-ray sensitivity and the only one outside of LEO with the capability of on-board transient localization. 

%------------------------------------------------------------------------

\section{POLAR-2}

\subsection{Mission Overview}

\begin{figure}  %13
\centering
\includegraphics[width=4.5in]{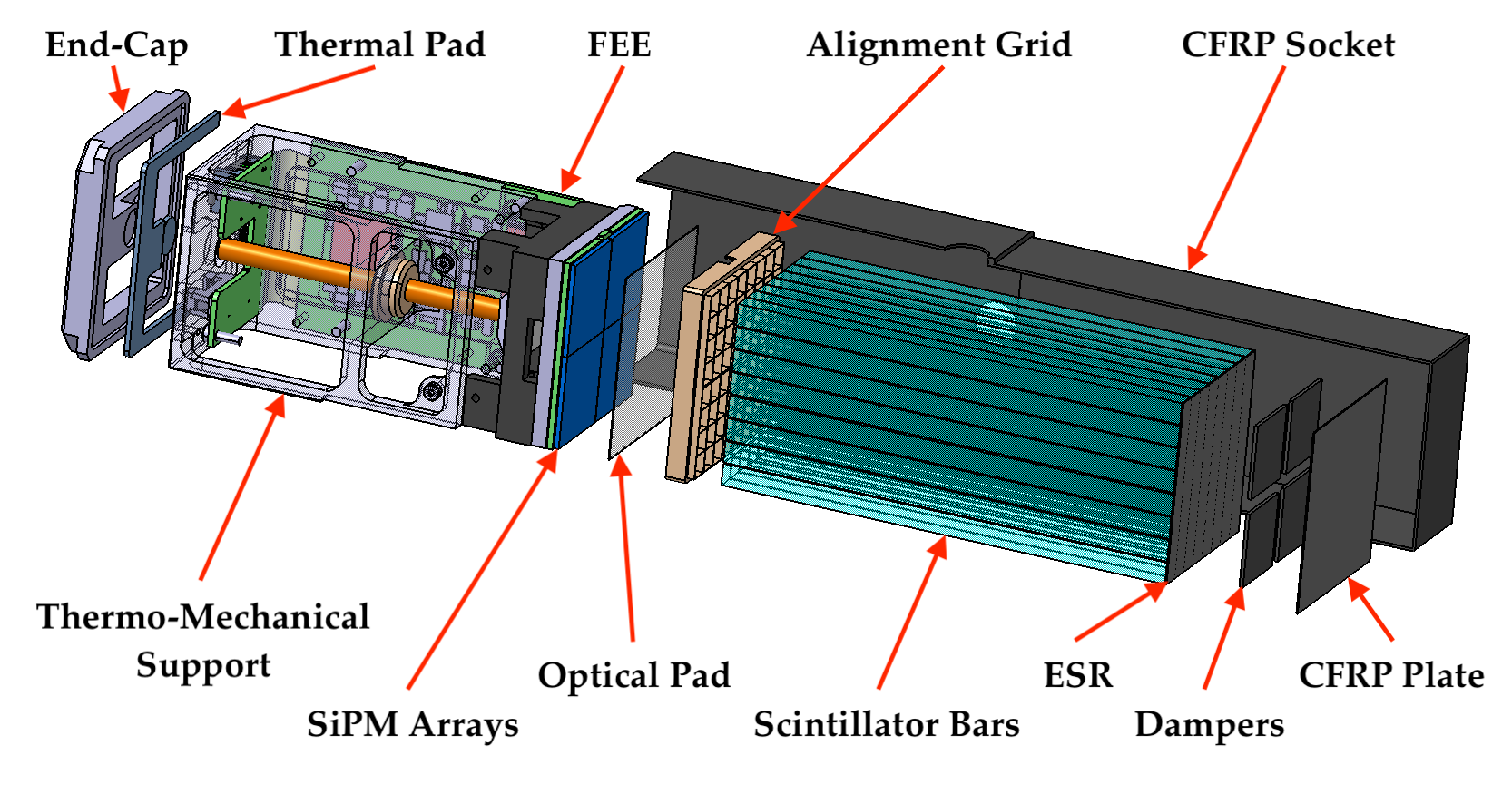}
\caption{An exploded view of one of the 100 POLAR-2 detector modules which make up the instrument. The detector measures the polarization of the incoming photons when these undergo Compton scattering in the segmented scintillator array. The scintillators are read out by a temperature-controlled Silicon photomultipliers (SiPM) array connected to its own front-end electronics. Taken with permission from \cite{NDA_Thesis}\label{fig:detector_module}.}
\end{figure}

POLAR-2\footnote{\url{https://www.unige.ch/dpnc/polar-2}} is a dedicated GRB polarimeter foreseen to be launched in 2027 towards the China Space Station (CSS). The detector is being developed by an international collaboration consisting of teams from Switzerland, Poland, Germany, and China. The project is currently ready for the production of the flight model after a prototype was successfully tested for physics performance and space qualification in 2023. POLAR-2 is a successor of the POLAR mission which detected 55 GRBs between October 2016 and April 2017 and performed polarization measurements in the energy range of 50--500\,keV of 14 of these \cite{Kole2020} as well as of the Crab pulsar \cite{Li2022}.

Although the measurement results of POLAR are the most constraining GRB polarization measurements to date, they are only able to constrain the polarization degree to be below $\approx40$\,\%. Therefore the results remain consistent with the majority of the existing theoretical predictions \cite{Gill2023}. This, along with the hint of an evolution of the polarization with time observed in 2 of the GRBs \cite{Kole2020}, indicates the need for a significantly more sensitive detector. For this purpose the POLAR-2 detector was initiated in 2017, followed by an approval for launch towards the CSS, through a United Nations Office for Outer Space Affairs call in 2019.

\subsection{Detector Design}

Compared to its predecessor, the POLAR-2 detector will be a factor 4 larger in size, thereby employing a total of 6400 plastic scintillator bars. These scintillators are read out in groups of 64, using segmented SiPM arrays connected to their own front-end electronics as indicated in Figure \ref{fig:detector_module}. As the GRB photons enter the detector they can undergo Compton scattering in the detector array, followed by photo-absorption in a second scintillator. Their azimuthal scattering angle can be constrained using the relative position of the two scintillators in which the photon interacted. As the photons will scatter preferentially perpendicular to their initial polarization, this measurement allows to determine the polarization of the incoming photon flux. 

While the increase in detector area provides an increase in effective area of a factor 4 compared to POLAR, further improvements to its design allow for additional improvements in its sensitivity. Particularly the switch from PMTs as used in POLAR, to SiPMs results in a significant increase in sensitivity at lower energies.

\begin{figure} %14
\centering
\includegraphics[width=3.7in]{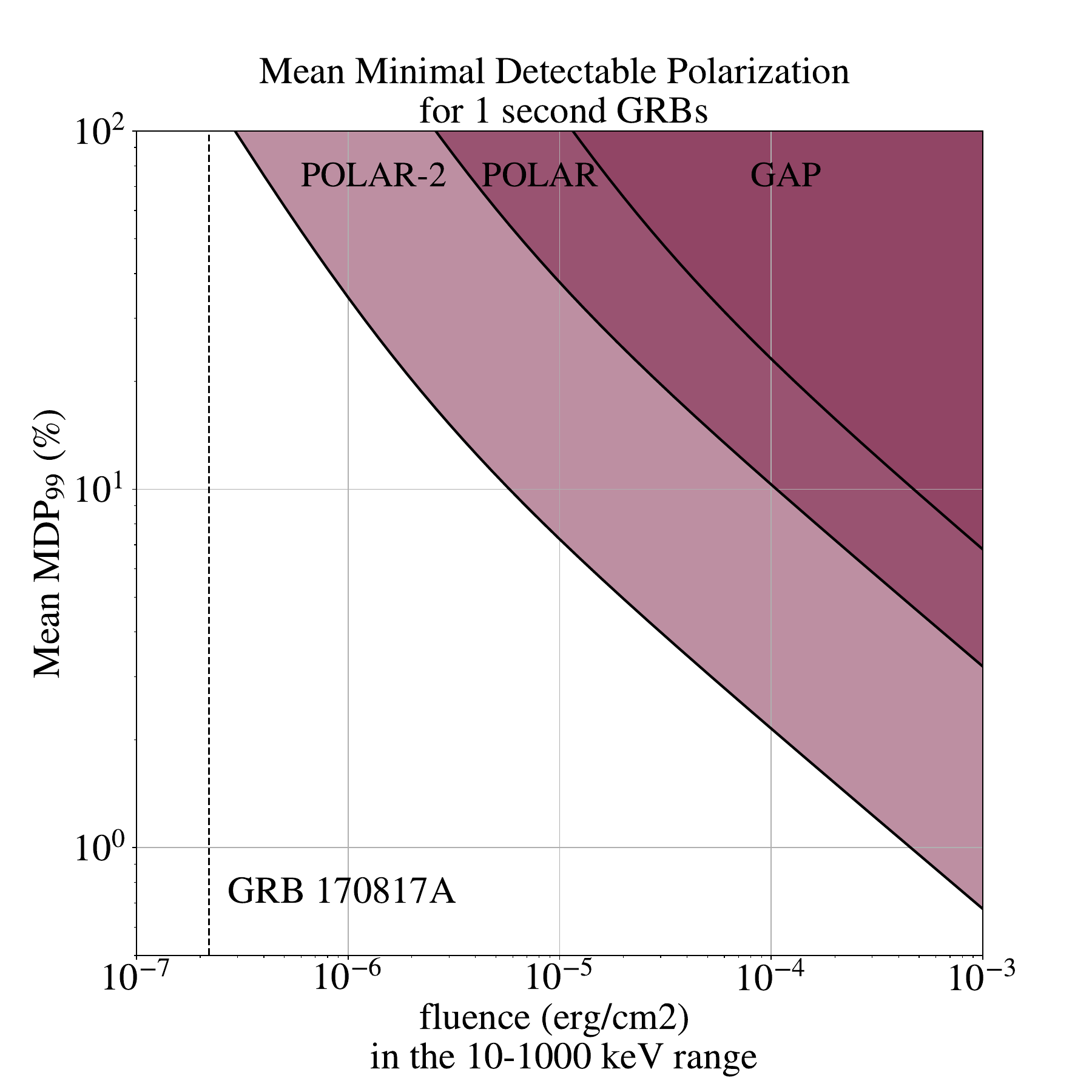}
\caption{The MDP (99\,\%) for 1\,s GRBs as a function of GRB fluence for the POLAR and POLAR-2 instruments. In addition, the sensitivity of GAP, the first dedicated gamma-ray polarimeter, is indicated. For reference, the fluence of the very weak GRB 170817A is indicated with a dotted line.\label{fig:MDP}}
\end{figure} 

\subsection{Expected Performance}

Thanks to all the design improvements POLAR-2 is approximately an order of magnitude more sensitive compared to POLAR as indicated in Figure \ref{fig:MDP}. As a result POLAR-2 will be able to perform constraining polarization measurements for GRBs with fluences as low as $10^{-6}$ erg\,cm$^{-2}$, while measurements able to constrain the polarization degree below 10\% will be possible for about 10--15 GRBs per year. 

The POLAR-2 detector will also contribute to transient alerts thanks to its large effective area which exceeds $2000\,\mathrm{cm^2}$. This large effective area, combined with continuous observations of half the sky and almost continuous communication to ground will allow POLAR-2 to send alerts within one minute from the onset of about 1 GRB every 2 days. As the instrument furthermore has access to a GPU onboard of the CSS, studies are currently ongoing on how to optimize GRB spectral and location information within such alerts \cite{HAGRID}.

%------------------------------------------------------------------------
\section{StarBurst}
\label{sec:starburst}
\subsection{Mission Overview}

The StarBurst Multimessenger Pioneer is a highly sensitive and wide field gamma-ray monitor designed to detect the prompt emission of gamma-ray bursts. StarBurst is designed as a SmallSat to be deployed to LEO as a secondary payload using the Evolved Expendable Launch Vehicle Secondary Payload Adapter (ESPA) Grande interface for a nominal 1-year mission starting in 2027 to coincide with  LIGO's scheduled 5th  observing run. StarBurst will utilize NASA's Tracking and Data Relay Satellites\footnote{\href{https://www.nasa.gov/mission/tracking-and-data-relay-satellites/}{https://www.nasa.gov/mission/tracking-and-data-relay-satellites/}} (TDRS) system to report possible electromagnetic counterparts to gravitational wave mergers with low-latency via the GCN system. StarBurst is among the first of NASA's new Pioneer class missions, intended to do compelling astrophysics science at a lower cost than missions in the Explorers Program. The Pioneers program provides opportunities for early-to-mid-career researchers to propose innovative experiments and lead space or suborbital science investigations for the first time.

\begin{figure}
\centering
\includegraphics[width=5.6in]{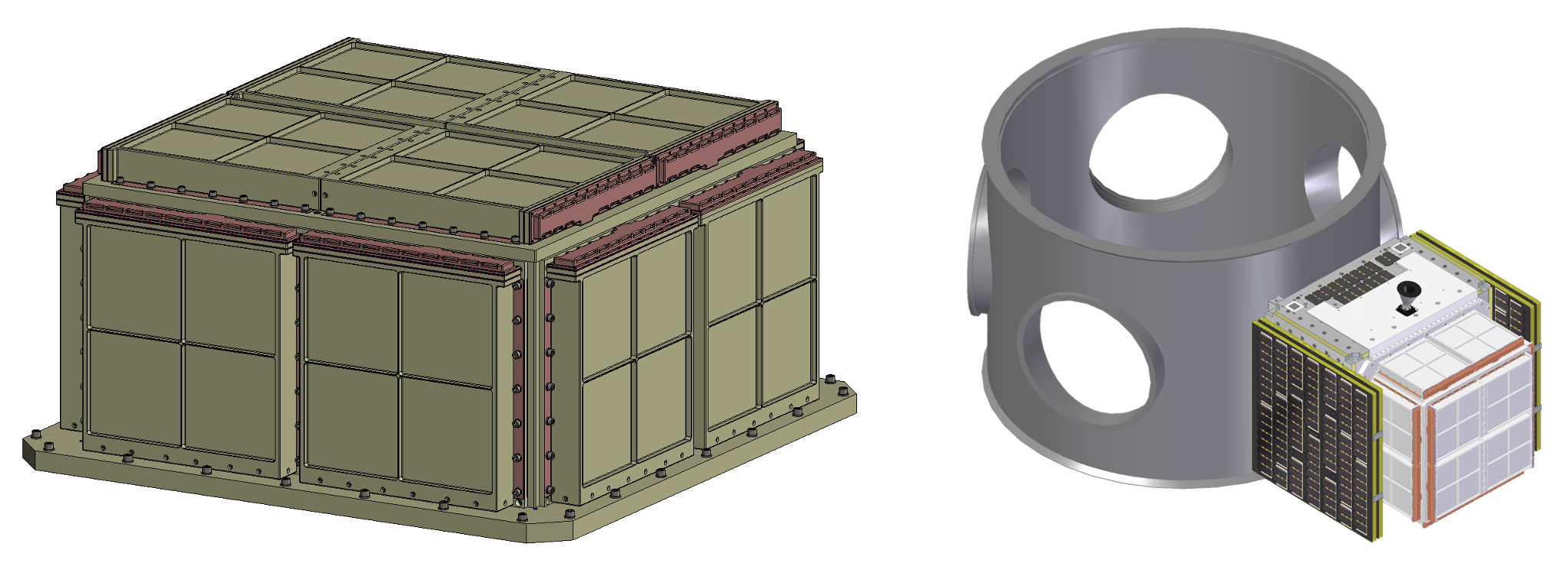}
\caption{\textbf{a}) The StarBurst instrument (left). \textbf{b}) The integrated StarBurst observatory shown mounted to an ESPA-Grande ring (right). \label{fig:starburst1}}
\end{figure} 

\subsection{Instrument Design}
StarBurst relies heavily on the heritage of the GBM, Glowbug, and BurstCube instruments, consisting of an array of 12 NaI(TI) scintillator detectors that utilize new, low mass and low voltage SiPMs to cover an energy range from 30\,keV to 2\,MeV. The StarBurst detectors are arranged to form a half cube, with two detectors mounted side-by-side to form four sides of the cube and four detectors comprising the top, as shown in Figure~\ref{fig:starburst1}. This configuration maximizes the available active surface area of the instrument given the volume constraints imposed on ESPA-Grande Rideshare Payloads and provides coverage of the entire unocculted sky. Each of the individual StarBurst detectors consists of a 24\,cm $\times$ 24\,cm $\times$ 1.6\,cm NaI(TI) scintillator read out by an array of SiPMs and enclosed in a housing with a thin (1-mm) aluminum window transparent down to 30\,keV and a beryllium-copper back shield to provide strong attenuation below $\sim200$\,keV. The scintillation light is read out by a 2 $\times$ 38 linear array of 6\,mm $\times$ 6\,mm J-Series SiPMs from ON Semiconductor (formerly SensL) optically coupled through an elastomeric silicone optical pad to a single edge of the NaI(Tl) crystal. Edge readout of the NaI(TI) exploits the planar geometry of the crystal to pipe scintillation light to the SiPM array. Each of the detectors has a dedicated bias voltage supply and Multi-Channel Analyzer (MCA) for independent operation. The MCA is a Commercial-off-the-Shelf (COTS) Bridgeport Instruments slimMorpho with an onboard 20\,MHz oscillator and flash ADC. 

\subsection{Expected Performance}

\begin{figure}
\centering
\includegraphics[width=5.5in]{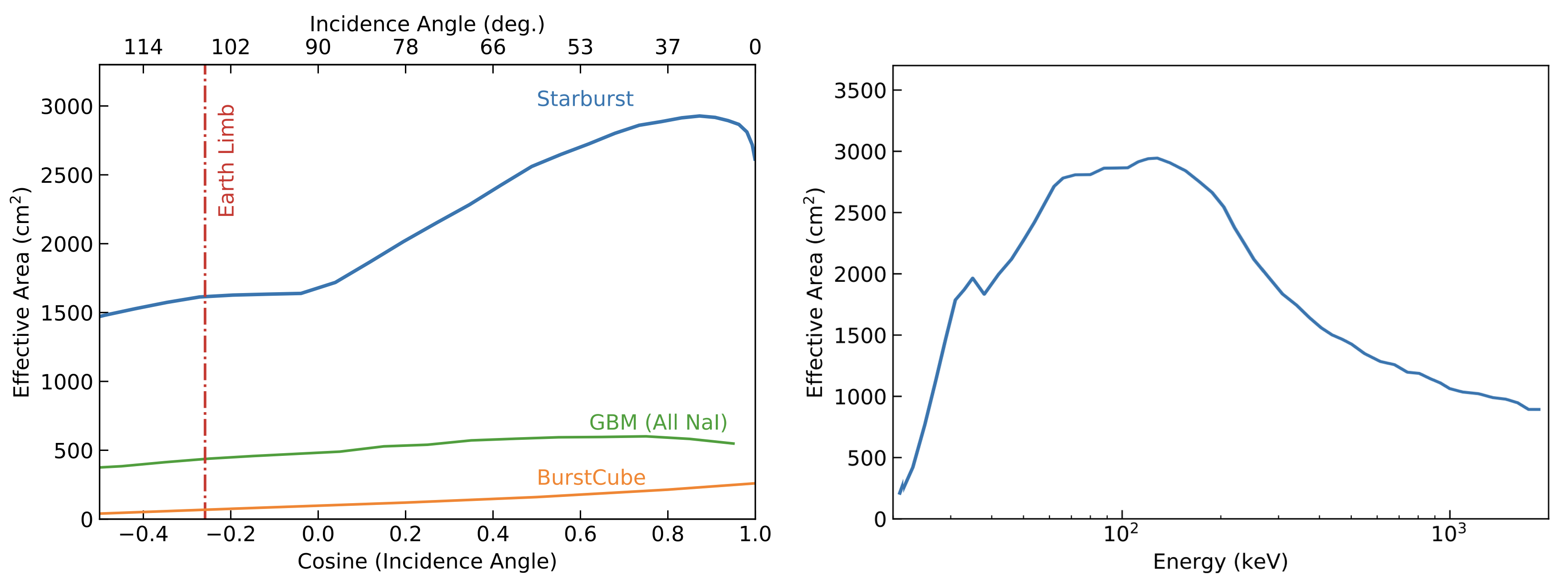}
\caption{\textbf{a}) The StarBurst effective area averaged over azimuth, as a function of angle from the instrument boresight (left). \textbf{a}) The StarBurst effective area as a function of energy (right). \label{fig:starburst2}}
\end{figure}

The scientific performance of the StarBurst instrument was assessed through Monte Carlo simulations using Geometry and Tracking 4 (GEANT4, \cite{AGOSTINELLI2003250}). The peak StarBurst effective area between 50 and 300\ keV, averaged over azimuth, is roughly 3000 cm$^{2}$, or roughly 500\,$\%$ that of GBM (Figure \ref{fig:starburst2}.  This provides an estimated detection efficiency of $>90\,\%$ for a 64-ms peak flux (photons\,cm$^{-2}$\,s$^{-1}$) of 0.9 for SGRBs ($< 2$\,s) and a 1024-ms peak flux of 0.25 for long GRBs ($> 2$\,s). Because StarBurst will have a similar field of view and duty cycle as GBM, it is estimated that StarBurst will detect 158 SGRBs per year, compared to the 40 and 8.6 SGRBs detected by GBM and Swift, respectively. StarBurst is estimated to achieve a localization uncertainty within 8$^{\circ}$ (1$\sigma$) for an SGRB with a 1 photon\,cm$^{-2}$\,s$^{-1}$ 64-ms peak flux. For SGRBs bursts with a 64-ms peak flux comparable to GRB 170817, or about 3 photons\,cm$^{-2}$\,s$^{-1}$, StarBurst is expected to achieve a localization uncertainty of $< 3^{\circ}$ (1\,$\sigma$). Employing the technique developed in \citet{Howell2019}, along with the StarBurst detection efficiency and the projected A+ (LHVKI) BNS detection efficiency provides an estimate joint GW-SGRB detection rate of roughly 7 joint GW-SGRBs detections per year.

%------------------------------------------------------------------------

\section{STROBE-X}
\label{sec:strobex}
\subsection{Mission Overview}

The Spectroscopic Time-Resolving Observatory for Broadband Energy X-rays\footnote{\href{https://strobe-x.org/}{https://strobe-x.org/}} 
(STROBE-X) is a mission proposed to the NASA call for an X-ray probe-class (\$1B PI-managed cost cap) mission responding to the recommendations of the 2020 Astrophysics Decadal Survey (Astro2020). It has a broad range of science goals with a focus on time-domain and transient events in the era of multi-wavelength and multi-messenger astronomy. It combines huge collecting area, high throughput on bright sources, broad energy coverage, and excellent spectral and temporal resolution in a single facility.  With its wide field of view, agile spacecraft, and low-latency communications, it would be a critical component of the NASA Time-Domain and Multi-Messenger (TDAMM) program. 

\subsection{Instrument Design}
\begin{figure}
\centering
\includegraphics[width=3.7in]{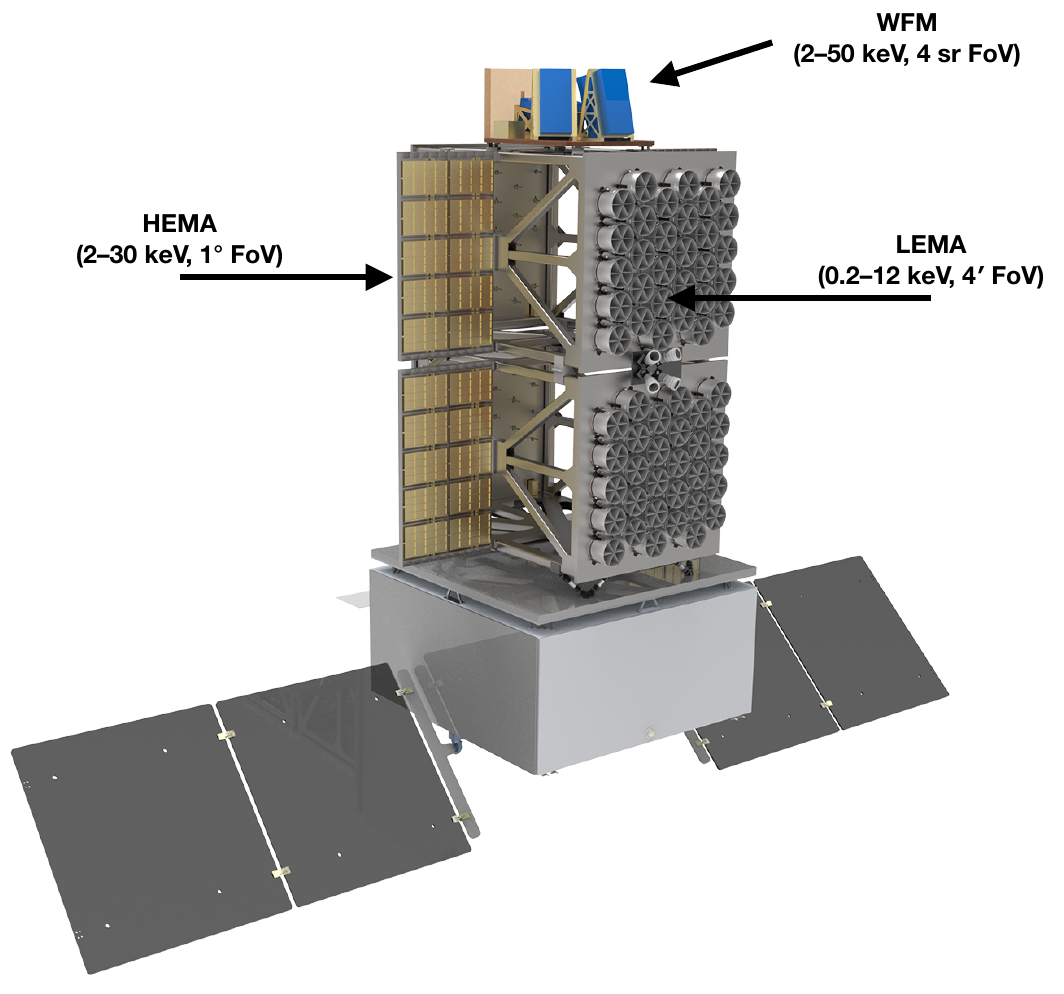}
\caption{Rendering of the STROBE-X spacecraft, showing the 3 science instruments. LEMA and HEMA are co-aligned narrow FoV instruments, while WFM has an instantaneous FoV of 1/3 of the sky.\label{fig:strobe-x}}
\end{figure} 

STROBE-X carries three instruments. The Low-Energy Modular Array (LEMA) covers the soft or low-energy band (0.2--12 keV) with an array of lightweight optics (3~m focal length) that concentrate incident photons onto small solid-state detectors with CCD-level (85--175\,eV) energy resolution, 100\,ns time resolution, and low background rates. This technology has been fully developed for NICER and will be scaled up to take advantage of the longer focal length of LEMA, which provides a factor of 8.5 improvement in effective area over NICER with over 1.6\,m$^2$. The High-Energy Modular Array (HEMA) covers the harder or higher-energy band (2--30 keV or beyond), with modules of Si drift detectors and micropore collimators originally developed for the European LOFT and eXTP mission concepts. HEMA provides a factor of 5.5 improvement in  effective area (3.4\,m$^2$) and $\sim 3$ in spectral resolution (200--300\,eV) over RXTE/PCA. The Wide-Field Monitor (WFM) comprises a set of coded-aperture cameras operating in the 2--50\,keV band that have a combined instantaneous field of view of 1/3 of the sky, with arcmin localization capability. It will act as a trigger for pointed observations of X-ray transients and will also provide high duty-cycle, high time-resolution, and high spectral-resolution monitoring of the dynamic X-ray sky. WFM will have 15 times the sensitivity of the RXTE All-Sky Monitor, enabling multi-wavelength and multi-messenger investigations with a large instantaneous field of view, down to a new, order-of-magnitude lower flux regime. Onboard processing will detect bursts in real time and provide notifications to the ground as well as triggering autonomous slews to get the pointed instrument on source in a few minutes.

\subsection{Expected Performance}

\begin{figure}
\centering
\includegraphics[width=3.7in]{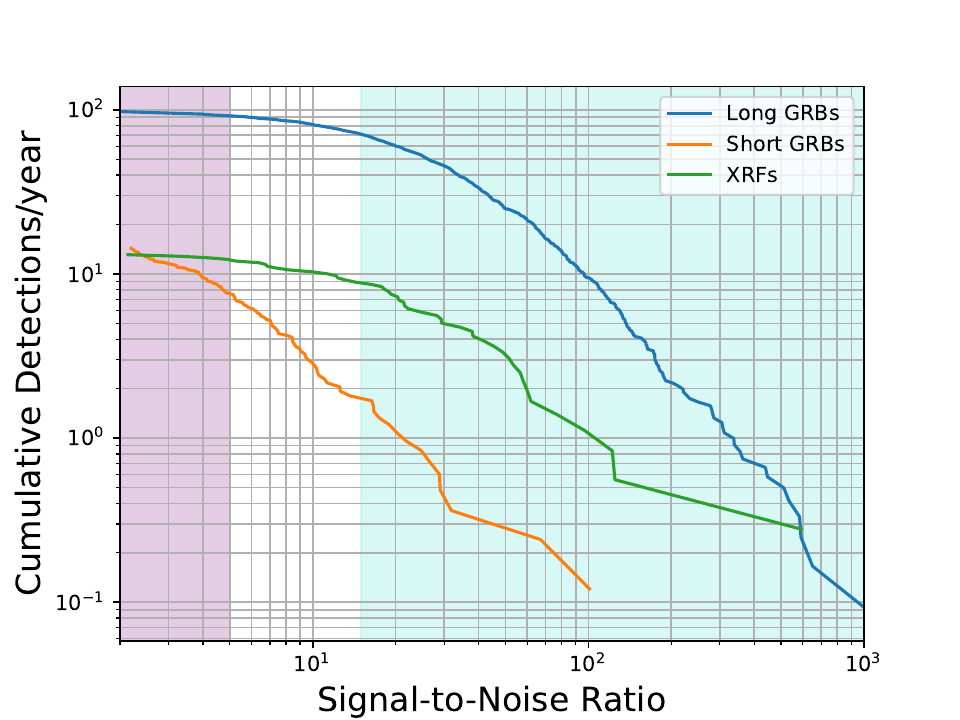}
\caption{The expected cumulative onboard detection rate of canonical gamma-ray bursts (GRBs) and X-ray flashes (XRFs) by the {\em STROBE-X}/WFM as estimated by folding observed GRB and XRF spectra through the WFM responses and accounting for the effective field-of-view of the WFM. It will detect $\sim 100$ long duration GRBs, $\sim 7$ short duration GRBs, and $\sim 12$ XRFs onboard per year. The onboard detection rate of long GRBs exceeds that of the {\em Swift}/BAT, while the short GRB detection rate is comparable.  A unique capability is the downlink of event data to the ground for the WFM, enabling sub-threshold searches to double the number of short GRB detections (purple shading).  The detection rate of XRFs exceeds that of previous instruments and is a particular science focus for the WFM.  The blue shading shows the region of signal-to-noise where high-fidelity spectroscopy can be performed in the prompt X-ray for these sources. \label{fig:strobe-x-grbs}}
\end{figure} 

STROBE-X will make major advances in several areas of GRB and multimessenger astronomy. The WFM will detect and localize $>5$ short GRBs per year and distribute the position accurate to 2 arcmin, brightness, and timing to the ground in $<5$ minutes. In some cases it will measure the redshift directly from the X-ray data (from the location of absorption edges in their spectra). STROBE-X will study the plateau emission of both short and long GRBs and provide unique diagnostics of whether the emission comes from millisecond magnetars or structured jets, revealing the nature of the central engines. 

STROBE-X will also be triggered by ground-based GW detections and get the pointed instruments on source within $<11.5$ minutes, given access to early afterglows with the tremendous collecting area and CCD-quality spectral resolution of LEMA.

The large grasp and softer response than many GRB missions will give STROBE-X access to unusual GRB phenomena that have not been well studied so far, including X-ray flashes (XRFs) and ultra-long GRBs (ULGRBs). Detecting and localizing $>10$ XRFs per year will increase the samples of these sources and break the degeneracies in the models for the source mechanisms.

%------------------------------------------------------------------------
\section{SVOM}

\subsection{Mission Overview}

\begin{figure}
\centering
\includegraphics[width=3.7in]{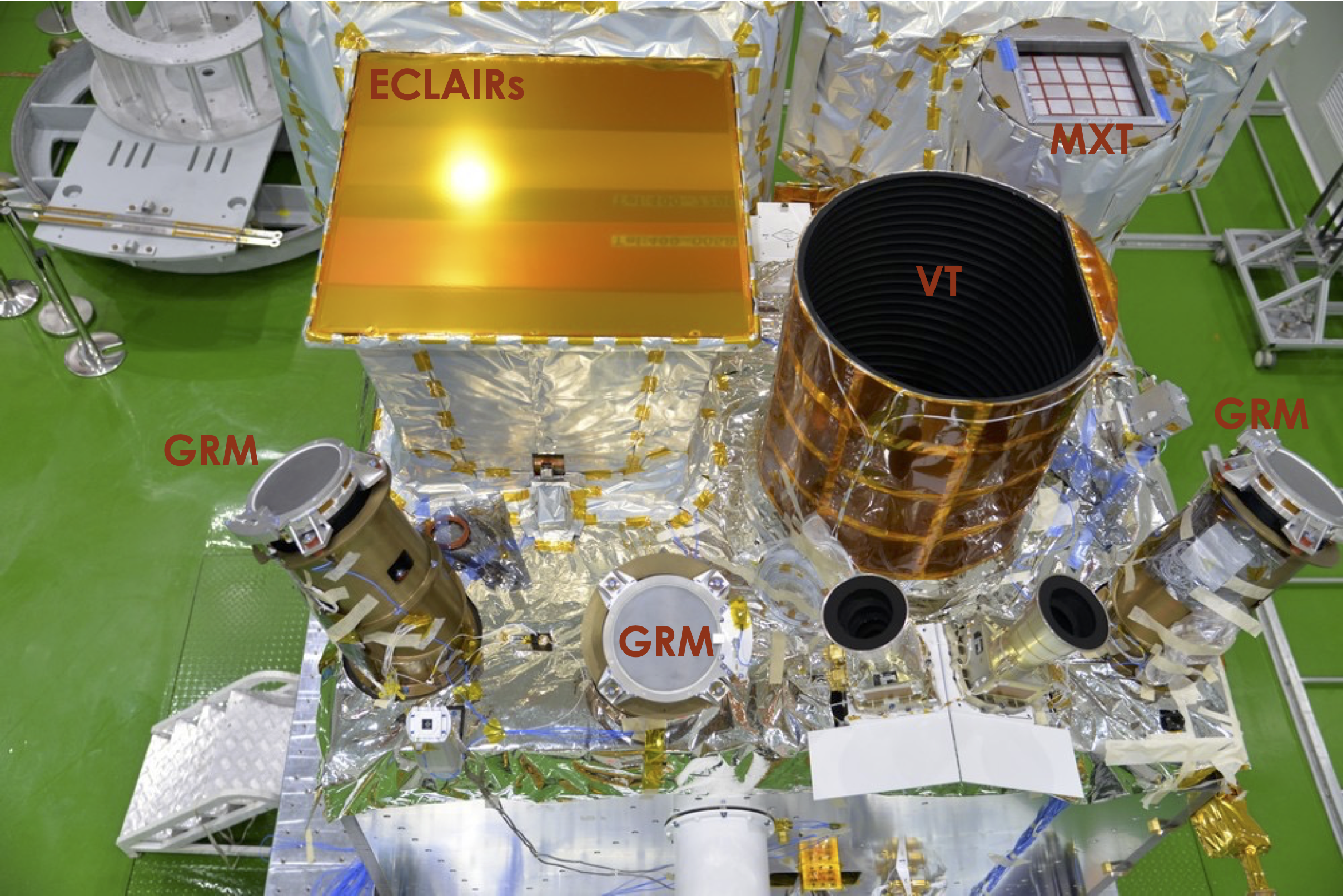}
\caption{The fully integrated SVOM payload. \label{fig:svom}}
\end{figure}

The Space based Variable astronomical Object Monitor (SVOM) mission is dedicated to gamma-ray burst studies and in general to Time Domain Astrophysics, including multi-messenger Science. It is optimized to detect and follow-up all type of GRBs, but particularly tailored for high-redshift GRBs thanks to its low energy triggering 
threshold around 4\,keV. One of the main goals of SVOM is to produce a complete catalogue of GRBs, including their redshift measurements. For this purpose SVOM will follow an anti-solar pointing strategy, and will avoid the galactic plane, in order to facilitate the observations from ground based observatories and robotic telescopes.  That is why SVOM is composed by a space segment, as well as a few ground based dedicated 
follow-up facilities. SVOM alerts will be promptly transmitted to the ground through a dedicated network of VHF antennas and through the Chinese Beidou inter-satellite communications link. The goal is that ground observers receive SVOM alerts less than 30\,s after the GRB is detected on board. The SVOM satellite will be launched from China around mid 2024 with a LM 2-C rocket and injected in a low Earth orbit (h$\sim$600\,km) with an inclination of about 30$^{\circ}$. It will carry four co-aligned instruments. Two instruments (ECLAIRs and GRM) are sensitive in the hard-X/soft gamma-ray energy range and have wide FoV, in order to monitor vast regions of the sky and detect gamma-ray transients. Two narrow FoV instruments (MXT and VT) will be used to follow-up and characterize the afterglow emission. The SVOM integrated payload is shown in Figure~\ref{fig:svom}.

\subsection{Instrument Design}

ECLAIRs is a coded-mask telescope, composed of a 54$\times$54\,cm$^{2}$ pseudo-random coded mask made of a 
Ti-Ta-Ti sandwich (10/0.6/10\,mm) placed 45.8\,cm above a pixellated detection plane made of 80$\times$80 CdTe crystals (4$\times$4$\times$1\,mm$^{3}$). Its FoV is about 2\,sr (89$^{\circ}\times$89$^{\circ}$) wide. ECLAIRs is sensitive in the 4\,keV--150\,keV energy range, and it comprises an on-board software to detect and localize (to better than 13\,arcmin) in near-real-time the GRBs that appear in its FoV. Once a new transient is detected, ECLAIRs issues an alert and requests the platform to slew so that the error box can be observed by the narrow-field instruments. 

ECLAIRs is complemented by the Gamma-Ray Monitor (GRM), a set of three 1.5\,cm thick NaI scintillators of 16\,cm in diameter, each one offset by 120$^{\circ}$ w.r.t. each other and with a combined FoV of $\sim$\,2.6\,sr. The GRM has poor localization capabilities, but it extends the SVOM spectral range up to about 5\,MeV, and increases the probability of simultaneous detection of short GRBs and GW alerts.

The Microchannel X-ray Telescope is a light ($<$\,42\,kg) and compact (focal length $\sim$\,1.15\,m) focusing X-ray telescope; its sensitivity below 1\,mCrab makes it the ideal instrument to detect, identify and localize down to the arcmin level X-ray afterglows of the SVOM GRBs.
Its optical design is based on a ‘‘Lobster-Eye'' grazing incidence X-ray optics, inspired by the vision of some crustacean decapods. It is composed of 25 square MPO plates of 40\,mm each arranged in a 5$\times$5 configuration. Although Lobster-Eye optics have been originally developed for large FoV telescopes (several tens of square degrees), the MXT optical design is optimized for a (relatively) small FoV of 58$\times$58 arcmin$^{2}$. The ``Lobser-Eye'' technique results in a peculiar point spread function (PSF), made by a central peak and two cross arms. The MXT optics is couples with a focal plane based on a pnCCD sensor, cooled at -65$^{\circ}$, sensitive in the 0.2--10\,keV energy band. MXT will localized GRB afterglows to better than the arc minute for the majority of them.

The Visible Telescope (VT) is a Ritchey-Chretien telescope with a 40\,cm diameter primary mirror. Its field of view is 26$\times$26 arcmin$^{2}$ wide, adapted to cover the ECLAIRs error box in most of the cases. It has two channels, a blue one (400--650\,nm) and a red one (650--1000\,nm), and a sensitivity limit of $M_{V}=22.5$ in 300\,s, allowing the detection $\sim$\,80\,\% of the ECLAIRs GRBs.
The main characteristics of the SVOM space segment are summarized in Table~\ref{tab:svom}.

   \begin{table}[ht]
    \centering
    \caption{Summary of the characteristics of the SVOM space instruments.}
    \renewcommand{\arraystretch}{1.25}
        \begin{tabular}{lllll}
            \hline %\hline
            & ECLAIRs & GRM & MXT & VT\\
            \hline
            Energy/Wavelength  & 4--150\,keV & 15--5000\,keV & 0.1--10\,keV & 650--1000\,nm\\
            Field of View &  2\,sr & 2.6\,sr (combined) &  58$^{\prime}$$\times $58$^{\prime}$ & 26$^{\prime}$$ \times$ 26$^{\prime}$ \\
            Localization accuracy  & $<$ 12$^{\prime} $& $<$20$^{\circ}$&$<$2$^{\prime}$ &$<$1$^{\prime\prime}$\\
            Expected GRBs year$^{-1}$ &60 & 90 & 50 & 40\\
            \hline
        \end{tabular}
    \label{tab:svom}
    \end{table}

The SVOM mission is also provided with a number of dedicated telescopes on ground. In particular here we mention:
\begin{itemize}
    \item the Ground-Based Wide Angle Cameras (GWACs), a set of 36 optical cameras with a combined FOV of 5400\,deg$^{2}$, located in Ali (China), whose goal is to catch the prompt optical emission for the ECLAIRs GRBs; 
    \item the Chinese Ground Follow-up Telescope (C-GFT), a robotic 1-m class telescope, with a 21$\times$21 arcmin$^{2}$ FOV, located in Xinglog (China) and sensitive in the 400-950\,nm wavelength range;
    \item the French Ground Follow-up Telescope (F-GFT, Colibri), a robotic 1-m class telescope, with a 26$\times$26 arc min$^{2}$ FOV, located in San Pedro Martir (Mexico), and with multi-band photometry capabilities over the 400-1700\,nm wavelength range. 
\end{itemize}

Other robotic telescopes will be part of the SVOM follow-up system, but they will not be fully dedicated to SVOM.

\subsection{Expected Performance}

The understanding of the GRB physics requires observations in the largest spectral domain and in the largest temporal interval, from the possible precursor up to the transition between the prompt and afterglow emissions. Simultaneous observation of the prompt GRB event in the gamma-ray, X-ray and visible bands, combined with narrow field observations of the afterglow in X-rays, visible and near infrared immediately after the beginning of the event will enable a better understanding of mechanisms at work in such events. New measurements (gravitational wave,
gamma-ray polarization, neutrinos) may also become possible in the future, and their impact on the physical understanding of the GRB phenomenon will be maximized if these measurements are made for bursts whose ``standard'' properties, including the distance, are well measured.
The SVOM mission is well adapted to these objectives. Compared to previous missions, it offers simultaneously (i) the capacity to trigger on all types of GRBs (especially on X-ray rich, and ultra-long ones); (ii) an excellent efficiency of the follow-up and the redshift measurement; (iii) a good spectral coverage of the prompt emission by ECLAIRs+GRM allowing a detailed modeling. For a significant fraction of SVOM GRBs, GWAC will provide in addition a measurement or an upper limit on the prompt optical emission;
(iv) a good temporal and spectral coverage of the prompt and afterglow emission thanks to MXT, VT, and the GFTs.

Concerning Multi-Messenger Astrophysics, SVOM with its ground and space instruments will offer a large and complementary follow-up capability
through ToOs. GWAC with its 5000 sq.\ deg.\ coverage can start the observation since the alert reception. The GFTs with their small FoV will confirm GWAC candidates and will be able to do follow-up for well localized events. To activate the satellite instruments, we will rely on a specific ToO program to send the observation program using S-band stations. This program guarantees less than 12 hours between the alert and the start of space observations (less can be expected for most cases) and can be activated around 20 times per year. From space, MXT and its 1 sq. deg.\ FoV will have the possibility to cover larger sky portion using a specific tiling procedure.

%------------------------------------------------------------------------
\begin{figure}[H]
\centering
\vspace{0.3truecm}
\includegraphics[width=5.4in]{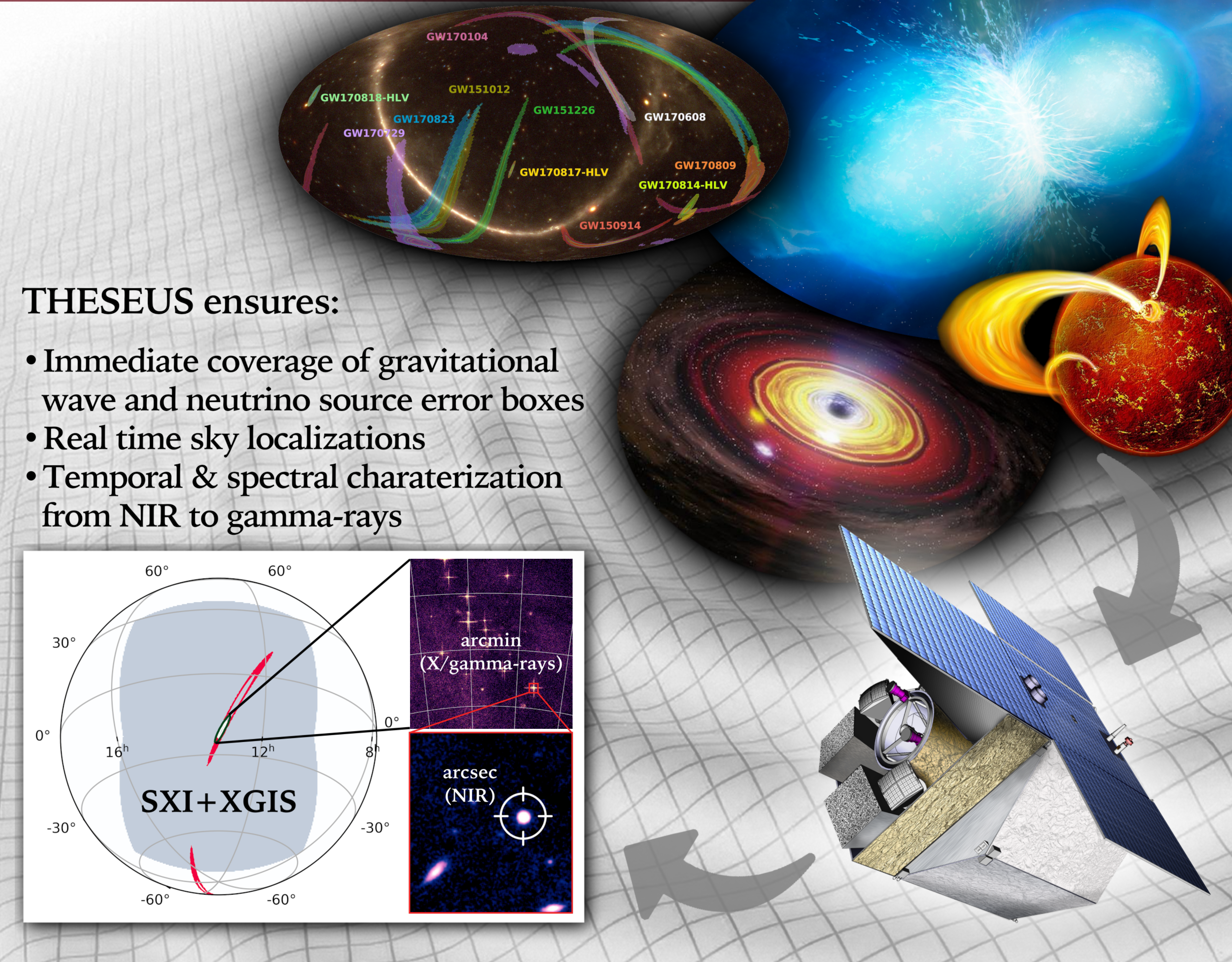}
\caption{Examples of THESEUS capabilities for multi-messenger and time-domain astrophysics.\label{fig:theseus1}}
\end{figure} 
\section{THESEUS}

\subsection{Mission Overview}

The Transient High Energy Sky and Early Universe Surveyor\footnote{\href{https://www.isdc.unige.ch/theseus/}{https://www.isdc.unige.ch/theseus/}} 
(THESEUS) mission concept aims to fully exploit GRBs for investigating the early Universe, around the epoch of re-ionization, and substantially advancing Multi-Messenger Astrophysics (see Figure~\ref{fig:theseus1}). THESEUS is planned to also simultaneously increase the discovery space of high energy transient phenomena and allowing tests of fundamental physics. The core science goals of THESEUS are summarized as follows:
\begin{itemize}
\item Investigating the first billion years of the Universe through high-redshift GRBs, thus shedding light on main open issues in modern cosmology, like: (i) the population of primordial low mass and luminosity galaxies; (ii) the drivers and evolution of cosmic re-ionization; (iii) the star formation rate (SFR) and metallicity evolution
up to the ''cosmic dawn'' and across Pop-III stars.
\item Providing a substantial advancement of multi-messenger and time-domain astrophysics by enabling the identification, accurate localisation and study of: (i) electromagnetic counterparts to sources of gravitational waves and neutrinos, which will be routinely detected in the mid-30’s by the second and third generation Gravitational Wave (GW) interferometers and future neutrino detectors; (ii) all kinds of GRBs and most classes of other X/gamma-ray transient sources.
\end{itemize}
The achievement of these scientific objectives will be possible by a mission concept including: (a) a set of innovative wide-field monitors with unprecedented combination of broad energy range, sensitivity, FOV and localization accuracy; (b) an on-board autonomous fast follow-up in optical/NIR, arcsec location and redshift measurement of detected GRB/transients. 

THESEUS has been selected twice by the European Space Agency (ESA) for a phase A study (in 2018 and 2022), aiming at demonstrating its technological and programmatic feasibility within the boundaries of a medium-size mission. THESEUS is currently undergoing its second phase A study, with a further selection step expected in 2026 and eventually a launch planned in 2037. Nominal scientific operations are being planned for four years but the lack of on-board consumable makes it feasible for the mission to extend operations in space well beyond the nominal lifetime. 

\begin{figure}
\centering
\includegraphics[width=3.1in]{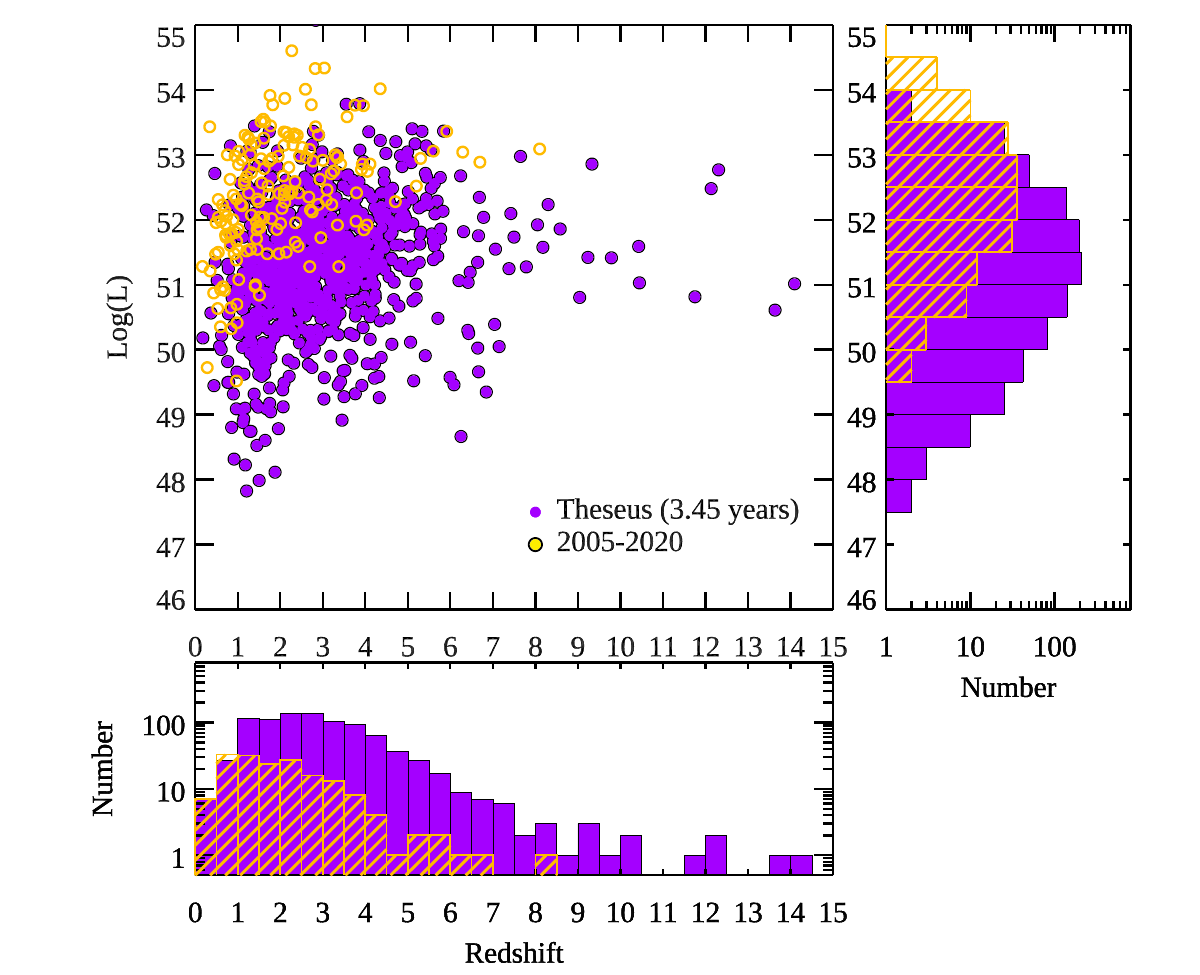}
\includegraphics[width=2.3in]{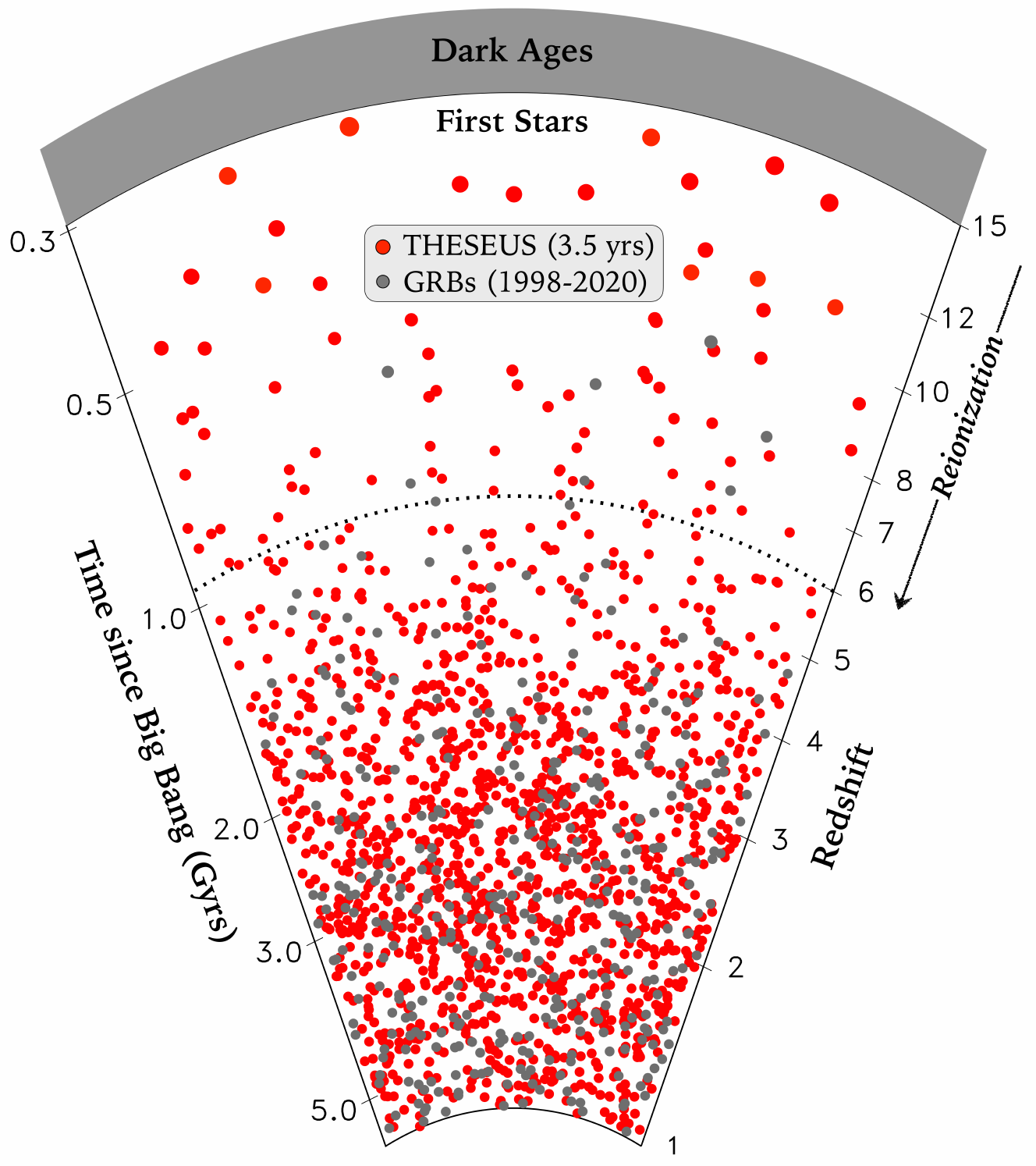}
\caption{{\it Left}: Distribution of long GRBs with redshift determination in the peak isotropic luminosity versus
redshift plane now (yellow points and hatched histogram) and after the nominal operation life of THESEUS (purple
points and full histogram). {\it Right:} A different version of the left side figure where the GRBs discovered by THESEUS and those detected up to 2020 are displayed in a cone representing the Cosmic evolution. \label{fig:theseus2}}
\end{figure}
\subsection{Instruments Design}

Three instruments are planned on-board THESEUS. The Soft X-ray Imager (SXI) uses lobster-eye wide-field ($\sim$0.5\,sr) focusing optics to increase the mission's sensitivity to fast transients in the 0.3--5\,keV energy band. The use of such optics provides uniform sensitivity across a very large field of view while maintaining arcminute localisation accuracy. The X and Gamma-ray Imaging Spectrometer (XGIS) is a GRB and transients monitor providing an unprecedented combination of exceptionally wide energy band (2\,keV--10\,MeV), imaging capabilities and location accuracy. The latter achieves $<15$\,arcmin up to 150\,keV over a FoV of 2$\pi$. The instrument is also characterized by an energy resolution of few hundreds eV at energies $<30$ keV and a time resolution of few $\mu$s over the whole energy band. Finally, the Infra-Red Telescope (IRT) is mainly conceived to detect, identify and measure the redshift of GRB afterglows detected by the SXI and the XGIS, especially those at high redshifts ($z>6$). The IRT is a 70\,cm Korsch telescope, optimized for an off-axis line of sight (LoS) of 0.884\,deg. The optical design will implement two separated FoV, one for photometry with a minimal size of 15$\times$15 arcmin (potentially extendable to 17$\times$20 arcmin), and one for spectroscopy of 2$\times$2\,arcmin. On the photometric field of view the IRT will be able to acquire images using five different filters (I, Z, Y, J and H) and on the spectroscopic field of view will provide moderate resolution (R$\sim$400) slit-less spectroscopy in the 0.8--1.6\,$\mu$m range. 

\subsection{Expected Performance}

THESEUS has two main core science objectives: the characterization of the physics of the high redshift universe and the electromagnetic characterization of gravitational wave transients (above all, the neutron star binary mergers). Long and short GRBs will be exploited to this scope. The first objective is instrumental to the understanding of the emergence of the first structures in the Universe, including the mass function of the high redshift galaxies, and the metal enrichment in the early
stages of the Universe (at the epoch of the reionization). The second objective has transformational potential, building upon the first ever gravitational wave multimessenger detection thus far, i.e.\ GW170817. THESEUS observations and discoveries in this field are expected to unveil the nature of ultra-dense matter, pin down the physics of relativistic jets, and help us understanding the nucleosynthesis of heavy elements (in turns probing fundamental aspects of general relativity and 
cosmology). We show in Figure~\ref{fig:theseus2} the total number of GRBs expected to be detected during the nominal lifetime of the mission (corresponding to 3.45\,yrs of scientific operations) compared to the total number of GRBs discovered from 2005 up to 2020. The transformational capabilities of THESEUS in these respects can well be appreciated from this figure.

\section{Conclusions}

In this paper we provided an overview of a number of missions, either close to beginning scientific operations or being planned for the coming/far future, that are expected to dramatically widen our capability of detecting and characterizing bright impulsive transient events of Astrophysical interest,  as (but not limited to) GRBs. The list of missions is not meant to be exhaustive, especially in view of the recent fast-growing interest of the international community in the fields of time-domain and multi-messenger Astrophysics which is driving many parallel developments in several countries around the globe. The missions presented here  have been or are being conceived based on largely overlapping science objectives which have, as main celestial targets, transient and fast variable sources (down to time scales as short as fractions of a second). Catching and studying these objects typically requires the simultaneous availability of large FoV X-/$\gamma$-ray instruments efficiently (i.e.\ with high duty cycles) monitoring the high energy sky and either narrower field instruments providing higher sensitivity measurements in complementary energy domains (from space as well as from the ground) or advanced detector technologies paving the way to poorly explored regions of the relevant parameter space (e.g., high energy polarimetry, simultaneous energy-time resolved spectroscopy, etc.). Given the largely different programmatic states of all summarized missions and the dynamic evolution of the design of at least those missions which are still in the early design and programmatic stage, providing an exhaustive comparison of their performance capabilities in different scientific fields is hardly achievable with sufficient confidence and thus beyond the scope of the present paper.   

At the time of writing, EP is the first of the described mission to come on-line, as the launch was successfully executed on 2024 January 9th and information on the first results are expected to be publicly available soon. SVOM is planned to be the second in line, with a launch planned for June 2024. As described in the previous sections, both missions are likely to boost the number of GRBs, as well as other high energy transients, to be discovered, characterized, and possibly followed-up in different energy domains during the next (at least) $\sim$5~years. The lifetime of both EP and SVOM will significantly overlap with the scientific runs of the current generation of GW detectors, as LIGO and VIRGO, possibly providing the discovery of electromagnetic counterparts of merging binary systems. A similar conclusion applies to the case of StarBurst, which launch is currently planned in 2027 and the 1-year expected duration of science operations shall match the so-called LIGO's scheduled fifth observing run. In 2027, also the installation of POLAR-2 onto the CSS is being planned and this should bring to the community a much deeper insight onto the polarization properties of GRBs, extending the outcomes of previous investigations in this domain from its predecessor, POLAR. 
Less certain is at present the path to the beginning of operations for a few more missions presented, eXTP, Gamow, HiZ-GUNDAM, LEAP, and MoonBeam. The advanced design state of all these missions could possibly bring them to space in between late 2020s to early 2030s, making them largely complementary and uniquely valuable for discoveries in the field of time-domain and multi-messenger Astrophysics to EP, SVOM, and StarBurst. STROBE-X and THESEUS are being planned for scientific operations not earlier than mid- to late-2030s. STROBE-X is competing with several other candidate missions in both the X- and IR domain within the NASA 2023 Probe mission call\footnote{\url{https://explorers.larc.nasa.gov/2023APPROBE/}.}, while THESEUS is competing against two further candidate missions for a launch opportunity in $\sim$2037 within the context of the ESA seventh call for medium-sized mission\footnote{\url{https://www.esa.int/Science_Exploration/Space_Science/Final_three_for_ESA_s_next_medium_science_mission}.}. Although the possible launch dates of STROBE-X and THESEUS are projected relatively far in the future, their timeline could interestingly match the planned observational runs from the 3rd generation of GW detectors, such as the Einstein Telescope \citep[ET; see, e.g.,][and references therein]{Punturo_2010, Maggiore_2020} and Cosmic Explorer \citep[CE; see, e.g.,][and references therein]{ce19}. The enhanced sensitivity of these instruments would greatly increase the number of possibly detected electromagnetic counterparts of GW sources, reaching up to a few tens per year as reported, e.g., in the case of THESEUS \citep{stratta22}.

%%%%%%%%%%%%%%%%%%%%%%%%%%%%%%%%%%%%%%%%%%
\vspace{6pt}

%%%%%%%%%%%%%%%%%%%%%%%%%%%%%%%%%%%%%%%%%%
\authorcontributions{E.\ Bozzo coordinated the overall work; 
W.\ Yuan and H.\ Sun have provided information on Einstein Probe; 
E.\ Bozzo, S.N.\ Zhang, M.\ Feroci, M.\ Hernanz, and A.\ Santangelo have provided information on eXTP;  
N.\ White, S.\ Guiriec T.-C.\ Chang, M. Seiffert, W. Baumgartner, C.\ Kouveliotou, and A.\ van der Horst have provided information on Gamow; 
D.\ Yonetoku, A.\ Doi, H.\ Matsuhara have provided information on HiZ-GUNDAM; 
M.L.\ McConnell, J.\ Gaskin, C.\ Wilson-Hodge and P.\ Veres have provided information on LEAP; 
C.\ Michelle Hui, C.\ Wilson-Hodge, A.\ Goldstein and P.\ Jenke have provided information on MoonBEAM; 
M.\ Kole, N.\ Produit and N.\ De Angelis have provided information on POLAR-2; 
D.\ Kocevski and Jon E.\ Grove have provided information on Starburst; 
P.\ Roming, C.\ Froning, P.S.\ Ray and T.\ Maccarone have provided information on STROBE-X; 
D.\ G\"otz, B.\ Cordier, and J.\ Wei  have provided information on SVOM;  
E.\ Bozzo, L.\ Amati, D.\ G\"otz, P.\ O'Brien and A.\ Santangelo have provided information on THESEUS. }

\funding{This research received no external funding.}

\dataavailability{This paper contains information about different missions either being planned for the future, or starting nominal operations soon. Further details on all missions considered can be found on the corresponding websites (if applicable) or on complementary material in the literature.}

\acknowledgments{We thank all referees for their constructive comments and suggestions. LA acknowledges support from the Italian Ministry of University and Research through grant PRIN MIUR 2020 - 2020KB33TP METE and INAF grant programme 2022.}

\conflictsofinterest{The authors declare no conflict of interest.}

%%%%%%%%%%%%%%%%%%%%%%%%%%%%%%%%%%%%%%%%%%
\begin{adjustwidth}{-\extralength}{0cm}
%\printendnotes[custom] % Un-comment to print a list of endnotes

\reftitle{References}

% Please provide either the correct journal abbreviation (e.g. according to the “List of Title Word Abbreviations” http://www.issn.org/services/online-services/access-to-the-ltwa/) or the full name of the journal.
% Citations and References in Supplementary files are permitted provided that they also appear in the reference list here. 

%=====================================
% References, variant A: external bibliography
%=====================================

\bibliography{allrefs.bib}

\PublishersNote{}
\end{adjustwidth}

\end{document}